\documentclass[aps,prd,showpacs,twocolumn,amssymb,superscriptaddress,
floatfix]{revtex4}

\usepackage{graphicx}
\usepackage{dcolumn}
\usepackage{bm}
\usepackage{color}%

\RequirePackage{cmap} 
\RequirePackage[TS1, T2A] {fontenc}
\RequirePackage[english] {babel}

\newcommand{\frat}[2]{\frac{\textstyle #1} {\textstyle #2}}

\newcommand{\vf}[1]{\mbox {\boldmath $ #1 $}}

\newcommand{\dmn}[2]{\mbox {$ #1 \! \cdot \! 10^{-2} \, $}}

\begin{document}

\title {Non-Abelian color fields from relativistic color charge
configurations \\ in the classical limit}

\author{W. Cassing}
\affiliation{Institute for Theoretical Physics, University of
Giessen, 35392 Giessen, Germany }

\author{V.V. Goloviznin}
\affiliation{Bogolyubov Institute for Theoretical Physics,
 National Academy of Sciences of Ukraine, 252143 Kiev, Ukraine}

 \author{S.V. Molodtsov }
\affiliation{Joint Institute for Nuclear Research, 141980 Dubna,
Moscow Region, Russia}
\affiliation{Institute of Theoretical and
Experimental Physics, 117218 Moscow, Russia}

\author{A.M.~Snigirev}
\affiliation{Skobeltsyn  Institute of Nuclear Physics, Lomonosov
Moscow State University, 119991 Moscow, Russia}

\author{V.D. Toneev}
\affiliation{Joint Institute for Nuclear Research, 141980 Dubna,
Moscow Region, Russia}

\author{V. Voronyuk}
\affiliation{Bogolyubov Institute for Theoretical Physics,
 National Academy of Sciences of Ukraine, 252143 Kiev, Ukraine}
\affiliation{Joint Institute for Nuclear Research, 141980 Dubna,
Moscow Region, Russia}

\author{G.M. Zinovjev}
\affiliation{Bogolyubov Institute for Theoretical Physics,
 National Academy of Sciences of Ukraine, 252143 Kiev, Ukraine}

\begin{abstract} We study the dynamics of color fields as generated by
simple configurations of relativistic particles with Abelian and
non-Abelian (SU(2)) charges in the classical limit. We find that
chromodynamic (non-Abelian) systems generally show Coulomb-like
features by analogy with electrodynamics. A peculiar feature in
the non-Abelian case is the additional strength of the
chromoelectric and chromomagnetic fields caused by the
contribution of changing the color charge. This change of color
SU(2) charges  results in a rotation of the color vector which
is getting very fast at close partonic distances. The presence
of this non-Abelian additional term in the chromoelectric and
chromomagnetic fields creates  a 'color charge glow',  which is
manifested as a distinct color wave disturbance arising due
to the finite distance at which the color interaction becomes
active. This situation may be relevant to the hadronization
phase in  ultrarelativistic heavy-ion collisions, where the
partonic state is governed by strong local color fluctuations.
 \end{abstract}

\pacs{25.75.-q, 24.85.+p, 12.38.Mh}

 \maketitle

\section{Introduction}
Ultra-relativistic nucleus-nucleus collisions at Relativistic
Heavy-Ion Collider (RHIC) or Large-Hadron-Collider (LHC) energies
provide an opportunity to explore strongly interacting QCD matter
in- and out-off equilibrium. The experiments at the RHIC and the
LHC have demonstrated that a stage of partonic matter is produced
in these reactions what is in approximate equilibrium for a
couple of fm/c \cite{RHIC1,RHIC2,RHIC3,RHIC4,LHC0}. Contrary to
early expectations, guided by  perturbative QCD (pQCD)
calculations, the medium shows a collective response of the
strongly interacting plasma (sQGP) with interaction rates beyond
those of high density hadronic matter \cite{xx1,xx2,xx3}. In fact,
viscous hydrodynamical calculations are successful in describing
the collective response of the partonic medium which is dominantly
reflected in the differential azimuthal angular distribution of
the final (observable) hadrons \cite{yy1,yy2,yy3}. These angular
distributions can be characterized by the Fourier coefficients
$v_n$ that specify the strength of the $n$-th harmonics in the
angular distributions. The observation of sizable uneven
coefficients $v_3, v_5$ showed the presence of initial-state
fluctuations which are due to random positions of the nucleons in
the targets and random inelastic hard interactions in the primary
phase of the nucleus-nucleus collisions  or due to color field
fluctuations in an early "glasma" phase. This comes about as
follows: the passage time of even central Au+Au (or Pb+Pb) at RHIC
energies is not longer than 0.2 fm/c at top RHIC energies and below
0.02 fm/c at the present LHC energies. During these short times the
wave functions of the nucleons in the target are frozen and even
the partonic wave functions of the individual nucleons are
approximately frozen so that configurations with high fluctuations
in the energy density show up in individual events
\cite{22,23,24,25,27}. In fact, when accounting for such initial
state fluctuations, all harmonics $v_n$ in the azimuthal angular
distributions can be described in event-by-event calculations
within the hydrodynamic framework \cite{Schenke} or within
transport approaches like the parton-hadron-string dynamics (PHSD)
theory \cite{Volodya12}.

The partonic medium (near an equilibrium state) is characterized
by a very low ratio of the shear viscosity to the entropy density
$\eta/s \approx$ 0.1--0.2 which is close to the lower bound of
$1/(4\pi)$ \cite{stari}. It is presently accepted that $\eta/s$
has a minimum as a function of temperature close to $T_c \approx$
160 MeV and increases fast in the hadronic phase, i.e., with
decreasing temperature \cite{vita1,vita2,vita3}. Furthermore, the
ratio of the bulk viscosity to entropy density is expected to have
a maximum close to $T_c$ \cite{zeta1,zeta2,zeta3}.

However, the (dissipative) hydrodynamic calculations start at
delayed times of 0.2 -- 1.0 fm/c from the initial impact of the
heavy ions since  approximate equilibrium has to be achieved by
presently unknown mechanisms \cite{Lappi}. Initial fluctuations in
the hydro calculations are usually imposed by independent Glauber
model simulations.
Initial conditions very similar to the Glauber model are included
by default in the PHSD transport approach that has been tested
successfully  so far for a variety of observables in
nucleus-nucleus collisions from lower super-proton-synchrotron
(SPS) up to LHC energies \cite{PHSD1,PHSD2,PHSD3,PHSD4}, including
electromagnetic probes such as $e^+ e^-$ or $\mu^+ \mu^-$ pairs
\cite{Linnyk}. The event-by-event PHSD calculations for heavy-ion
reactions from RHIC to LHC energies show that apart from local
fluctuations in the energy density also sizable fluctuations in
"color" occur although the total system is always color neutral.
Furthermore, the formation of color dipoles is observed in the
microscopic transport calculations, in particular in the
hadronization phase at rather low parton densities, which implies
that color forces might play an important role  during the early
non-equilibrium stage (cf. Ref. \cite{Mishustin}) as well as in
the dynamics of hadronization.

 The dynamical evolution of such systems  depends on the initial
conditions for the above-noted string-like mechanism and
additionally on the parton distribution function in the nuclei
prior to the collision. High-energy heavy-ion collisions initially
release a large number of gluons from their wave functions. In
fact, the wave function of a hadron boosted close to
 the light cone is densely packed with gluons so that they
may “overlap” \ leading to  saturation in the gluon density due to
nonlinear gluon interactions~\cite{GLR83}. Therefore, at high
collision energies the colliding hadrons can be viewed as high-density
gluon fields. This dense system is nowadays referred to as  a color
glass condensate (CGC) (cf. the review articles~\cite{GIJV10}).

The problem of computing the distribution functions for gluons at
very small Bjorken $x$ has been a task for decades~\cite{GIJV10}. The gluon
and quark distribution functions are computable in perturbation
theory at large values of the Bjorken $x$, but at small $x$ the
precise computation encounters  much uncertainty.  The
perturbative paradigm of the low $x$ gluon physics is the Balitsky-Fadin-Kuraev-Lipatov (BFKL)
equation~\cite{BFKL}, which, however, is linear and ignores mutual
interactions of the gluons.  Thus, nonperturbative calculations
are needed.  The nuclear structure function for large parton
density at small $x$, where classical methods are applicable, is
rather successfully described by the McLerran-Venugopalan
model~\cite{McLV94}  within a classical effective field theory
approach, where the degrees of freedom are static color sources in
the hadron at large $x$, coupled to the dynamical gluon fields at
small $x$. It  recently has been shown that renormalization group
generalization of this effective action can improve this approach.
The renormalization group equations, derived by requiring that
observables are independent of the separation in $x$ between
sources and fields, lead to an infinite hierarchy of evolution
equations in $x$. With appropriate initial conditions, the
solutions of this
Jalilian-Marian,Iancu,McLerran, Weigert, Leonidov,and Kovner (JIMWLK)
hierarchy~\cite{JIMWLK}  allow one to compute
a wide range of multiparticle final states in deeply inelastic
scattering or/and hadronic collisions and to reproduce several key
results in small $x$ QCD.

If a single heavy nucleus (such as Au or Pb) is considered within
the McLerran-Venugopalan model~\cite{McLV94}, an ultrareletivistic
collision of two such nuclei at the very early times can be
treated as a collision of CGC sheets which are sources of the
Yang-Mills radiation. The equations that describe the color field
evolution  were applied to RHIC energy collisions of nuclei
and deuteron-Au reactions~\cite{phenom}. This intermediate matter
is highly coherent and responsible for the transition from CGC
to the quark-gluon plasma where  fast thermalization might be reached
due to field instabilities. This intermediate state is called the
"glasma" which is gluon rich and low in the quark/antiquark
density.

One should emphasize that the CGC picture refers to the very early
interaction stage of ultrarelativistic collisions. To validate
this picture -- by comparing its predictions with experiment -- the
glasma model has to be supplemented by hydrodynamic or kinetic
models that transport the energy density and its fluctuations to
the final hadronic spectra. Both model types as well as the
matching procedure, connecting the two different descriptions of
the system evolution, introduce additional uncertainties which
prevent unambiguous conclusions, especially in the case of
nucleus-nucleus  collisions. In particular, the flow harmonics,
being sensitive to the early stage of the reaction, have been
compared using the CGC and Glauber initial conditions. Both model
versions, being progressively developing~\cite{Al11,GJS13}, give
very close results with a small advantage in favor of one or
another initial condition in different studies. The first LHC data on
the bulk particle production in Pb+Pb collisions are in good agreement
with improved CGC expectations but they are also compatible with
Monte Carlo event generators. Both approaches have in common that
they include strong coherence effects. Exhaustive analysis of
forthcoming more differential observables is needed to better
discriminate between the models (cf.~\cite{81,GJS13a}).

Our present study is, furthermore, motivated by recent
investigations of peripheral ultrarelativistic heavy-ion
collisions at the RHIC and the LHC, for which a large electric
charge $eZ$ of the colliding nuclei leads to the generation of
intense electric and magnetic fields during the passage time of
the charged "spectators"\ , as discussed in Ref.~\cite{Ton}. It is
speculated that a very strong electromagnetic field of short
duration essentially in the pre-equilibrium phase might have an
important impact on particle production~\cite{Tu13,Bl13}.

Future ALICE measurements are promising to provide an answer
to these questions and, possibly, to strengthen the standard
approach to the saturation physics~\cite{GIJV10}  treating the
gluon fields and the corresponding scattering cross sections
classically (rapidity-independent). Such classical field dynamics
call for a transport formulation and the development of an
extended transport code for ultrarelativistic collision
processes, including the color dynamics of the gluon fields
(beyond the processes implemented, e.g., in PHSD).

In order to proceed to a solution of this task, we are going here
to give  practical estimates for the space-time dependence of
classical non-Abelian fields which (in line with PHSD) are
generated by colored quarks in relativistic heavy-ion
collisions. In this respect we will develop an instrumental
approximation for configurations of two color charges, a color
charge and a color dipole and two color dipoles which are moving
along a straight line towards each other. For our initial study of
non-Abelian fields we simplify the problem to the SU(2)-color
group. We recall that such color configurations are just the
elementary basic examples in ultrarelativistic heavy-ion
collisions in the existing phenomenological
models~\cite{PHSD1,Kapusta}. The classical character of the
non-Abelian field means, as usual, that the field operators are
replaced by their average values for the quantum state, and the
off-diagonal elements are neglected. Quarks are treated as
classical point-like massive particles possessing a color charge.
We will estimate the characteristic values of the classical gluon
field strength but are not interested in the parton (gluon)
distribution functions themselves. More precisely, we will specify
the features characteristic of the non-Abelian nature of fields in
comparison to solutions for Abelian fields for the same coupling
strength.

The actual layout of our study is as follows: In Sec. II, we
first recall the results from classical (Abelian) electrodynamics
considering the retarded electromagnetic field created by a moving
source. In Sect. III, the particularities of the non-Abelian color
field produced by a color source are discussed on the basis of an
approximate solution of the classical Yang-Mills equations. The
application of this technique to particular cases of two color
charges, a charge-color dipole system and two dipoles moving in
the opposite direction is presented in Secs VI, V, and VI,
respectively. Our conclusions are given in Sec. VII. Some more
technical details for the color interactions and the construction
of approximate solutions of the classical Yang-Mills equations are
shifted to the Appendix.


\section{The Field of a Relativistic Abelian Charge}
In this section, we recall the results from classical electrodynamics:
The field of a point-like charge propagating along the trajectory
$\vf{r}(t)$ is described by the (retarded) Li\'enard-Wiechert
potential at the observation point $\vf{r}_0$
\begin{equation}
\label{1}
\varphi=\frat{1}{4 \pi}~\left[\frat{e}{R-\vf{v} \vf{R}}\right]_{t'},~~~
\vf{A}= \frat{1}{4 \pi}~\left[\frat{e~\vf{v}}{R-\vf{v}\vf{R}}
\right]_{t'}~.
\end{equation}
Throughout the paper we stick to the standard system of units with
speed of light $c=1$, dimensionless electrodynamic $e$ and
non-Abelian $g$ charges, i.e., have the proper factors of $\hbar
c$ to keep the proper dimensions. Then
characteristic distances in the problem are of an order of one
Fermi and the potentials and field strengths are measured in units
of $m_\pi$ and $m_\pi^2$, respectively, where $m_\pi$ is the
$\pi$-meson rest mass. In Eq. (\ref{1}), $\varphi$ is the zeroth
component of the potential and $\vf{A}$ is its vector component,
$\vf{v}$ is the particle velocity at some retarded time $t'$ which
is determined by the distance between the observation point and
the particle $R=|\vf{R}|$ where $\vf{R}=\vf{r}_0-\vf{r}(t')$ is
the radius-vector from the charge position to the observation
point $\vf{r}_0$. It is of importance to note that the relation
between the retarded and laboratory time $t$ is
\begin{equation}
\label{2}
R=t-t'~,
\end{equation}
(although the notation $R'$ would be more suited). The electric and
magnetic fields are then given by
\begin{equation}
\label{3}
\vf{E}=-\frat{\partial\vf{A}}{\partial t}-\vf{\nabla}\varphi~,~~
\vf{H}=\vf{\nabla}\times \vf{A}~.
\end{equation}
Using  Eq. (\ref{2}) we get
\begin{eqnarray}
\label{4}
\frat{\partial \vf{A}}{\partial t}&=&\frat{\partial
\vf{A}}{\partial t'} \frat{\partial t'}{\partial t}~,~~
\vf{\nabla}\times\vf{A}=\vf{\nabla}\times\varphi\vf{v}~,  \\
\frat{\partial t'}{\partial t}&=&\frat{1}{1-\vf{v}\vf{n}},~
\vf{\nabla}t'=-\frat{\vf{n}}{1-\vf{v}\vf{n}},~
\vf{\nabla}R=\frat{\vf{n}}{1-\vf{v}\vf{n}}~, \nonumber \\
\vf{\nabla}(\vf{v} \vf{R})&=&\vf{v}-\frat{\vf{n}}{1-\vf{v}\vf{n}}
(\dot{\vf{v}}\vf{R}-\vf{v}^2)~~~\vf{\nabla} \times\vf{v} \nonumber
\\&=&\vf{\nabla}t'\times \frat{\partial \vf{v}} {\partial
t'}=-\frat{\vf{n}}{1-\vf{v}\vf{n}}\times\dot{\vf{v}}\nonumber
\end{eqnarray}
with the unit vector $\vf{n}=\vf{R}/R$ while
$\dot{\vf{v}}=\partial \vf{v}/\partial t'$ is the particle
acceleration retarded in time (see Ref.~\cite{RT}). Combining
these relations and using the potentials from Eq. (\ref{1}) we get
the following results for the electric and magnetic fields:
\begin{eqnarray}
\label{5}
\vf{E} &=& \frat{1}{4 \pi}~\left[\frat{e}{R^2}
\frat{(1-\vf{v}^2)(\vf{n}-\vf{v})}{(1-\vf{v}\vf{n})^3} +
\frat{e}{R}\frat{\vf{n}\times (\vf{n}-\vf{v}) \times\dot{\vf{v}}}
{(1-\vf{v}\vf{n})^3}\right]_{t'},
\nonumber \\[-.1cm]
\\[-.25cm]
 \vf{H} &=& \frat{1}{4 \pi}~\left[-\frat{e}{R^2}\frat{\vf{n}\times
\vf{v}} {(1-\vf{v}\vf{n})^3}(1-\vf{v}^2 +\dot{\vf{v}}\vf{R}) -
\frat{e}{R}\vf{n}\times \dot{\vf{v}} \right]_{t'} \nonumber \\
&=& \vf{n}\times \vf{E}~. \nonumber
\end{eqnarray}
In fact the formulas (convenient for practical purposes) express the
electric and
magnetic fields as a function of the particle trajectory (at
$\dot{\vf{v}}=\vf{0}$) at the current time
$\vf{r}=\vf{r}_0-\vf{r}(t)$ in the form
\begin{equation}
\label{7}
\vf{E}=\frat{1}{4\pi}~\frat{e~\vf{r}(1-\vf{v}^2)}
{\left[(\vf{r} \vf{v})^2+
\vf{r}^2(1-\vf{v}^2)\right]^{3/2}}~,~~~~\vf{H}=\vf{v}\times\vf{E}~.
\end{equation}
We also recall that the potentials  can then be written as
\begin{eqnarray}
\label{10} \varphi&=&\frat{1}{4\pi}~\frat{e}{[(\vf{r}\vf{v})^2
+\vf{r}^2(1-\vf{v}^2)]^{1/2}},
 \nonumber \\[- .21 cm] \\[- .2 cm]
~~\vf{A}&=&\frat{1}{4\pi}~\frat{e~\vf{v}}{[(\vf{r}\vf{v})^2 +
\vf{r}^2(1-\vf{v}^2)]^{1/2}}. \nonumber
\end{eqnarray}

\section{The Field Of Relativistic SU(2) Charges (Classical Yang--Mills Fields)}
In order to keep our approach transparent, we consider
here  SU(2) non-Abelian fields only. In the classical
approximation they are $c$-number functions, which are the
solutions of the classical Yang-Mills equations. We analyze the
QCD Lagrangian of the form
\begin{equation}
\label{11}
{\cal L} = -\frat{1}{4} \widetilde G^{\mu \nu}
\widetilde G_{\mu \nu}-\widetilde j^\mu \widetilde A_\mu~,
\end{equation}
where the color vector $\widetilde
A_\mu=(A^1_\mu,A^2_\mu,A^3_\mu$) represents a triplet of the
Yang-Mills fields of different colors ("isospin"), $\widetilde
j^\mu $ is the current density of external color sources, and
$\widetilde G_{\mu\nu}=\partial_\mu \widetilde
A_\nu-\partial_\nu\widetilde A_\mu +g\widetilde
A_\mu\times\widetilde A_\nu $ is the  gluon field tensor with
the covariant derivative acting as $\widetilde {D^\mu
f}=\partial^\mu\widetilde f+g\widetilde A^\mu\times\widetilde f$.
The product sign $\times$  corresponds to the vector product in the color
space. The classical equations of motion, as known, then read
\begin{equation}
\label{12}
\widetilde {D^{\mu}G}_{\mu \nu}=\widetilde j_\nu~.
\end{equation}
The most significant difference between these equations and the
electrodynamic equations is the compatibility conditions of the system
of Eqs.~(\ref{12})
\begin{equation}
\label{13}
\widetilde {D^{\mu} j}_{\mu}=0~
\end{equation}
which generally implies that the color vector charge is not conserved
in a sense similar to electrodynamics. Then as
a suitable solution of the Yang- Mills equations for a single
particle with constant color charge $\widetilde C$ one may take
the potentials of the form
\begin{equation}
\label{14}
\widetilde \varphi=\frat{1}{4\pi}~
\left[\frat{\widetilde C} {R-\vf{v} \vf{R}} \right]_{t '},~~~
\widetilde {\vf{A}}=\frat{1}{4\pi}~\left[\frat{\widetilde C~\vf{v}}
{R-\vf{v}\vf{R}}\right]_{t'}~.
\end{equation}
Similarly to the electromagnetic case, discussed above in Sec.
II, $\vf{R}=\vf{r}_0-\vf{r}(t')$ is the radius-vector directed
from the charge position to the observation point $\vf{r}_0$. In
electrodynamics one can easily verify the uniqueness of the
solutions (\ref{1}) for a pointlike charge (taking into account
the well-known problem of the singularity associated with the
self-interaction~\cite{Rohr}) because the superposition principle
for solutions remains valid. Furthermore, the general form of the
retarded solution is well known. In the non-Abelian case, we cannot
benefit from an analysis of the general Yang-Mills solutions but
one can construct  approximate solutions with properties similar
to the solutions of electrodynamics. As mentioned above, here we
focus on the analysis of some simple examples important for
practical applications treating the relativistic color objects
propagating along a straight line.

As an approximate solution, we consider a superposition of the
Li\'enard-Wiechert potentials (\ref{14}) in which the vector of
the particle color charge can change in time  and should be taken
at the retarded time,  which results exactly from the general form
of the retarded solution as
\begin{eqnarray}
\label{21} &&\widetilde \varphi({\vf r},t)=\frat{1}{4\pi}~\int
\frac{\widetilde \rho({\vf r}',t')~ \delta(t-t'-|{\vf r}-{\vf
r}'|)}{|{\vf r}-{\vf r}'|}~{d^3r'} dt'~,
\nonumber\\[-.2cm]
\\[-.25cm]
&&\widetilde {\vf A}({\vf r},t)=\frat{1}{4\pi}~\int
\frac{\widetilde {\vf j}({\vf r}',t') ~\delta(t-t'-|{\vf r}-{\vf
r}'|)}{|{\vf r}-{\vf r}'|}~{d^3r'} dt'~.\nonumber
\end{eqnarray}
As an example, we consider a simple approximation of the covariant
four-current with the delta function for a pointlike particle
\begin{equation}
\label{22} \widetilde j_ \mu ({\vf r}',t')= (\widetilde C ~ \delta
({\vf {\vf r}'} - {\vf r} (t)), \widetilde C ~ {\vf v} (t) ~
\delta ({\vf r}' - {\vf r} (t))) ~,
\end{equation}
where ${\vf v}(t)=\dot{\vf r}(t)$ is the particle velocity. In
this context the compatibility condition (\ref{13}) for a
pointlike charge becomes
\begin{equation}
\label{23}
\dot {\widetilde C}=g~\left[\widetilde \varphi(t,{\vf r})-{\vf v}~
\widetilde {\vf A}(t,{\vf r})\right]\times\widetilde C~.
\end{equation}
From these equations it is easy to see that the modulus of the
color charge vector remains constant.
 One should note that in this approximation of  color
field sources as pointlike charges the evolution of a separate
"parton" \ is described by Eq.~(\ref{23}) due to the continuity equation.
Moreover, in the developed consideration, as noted in the
Introduction of Ref.~\cite{JIMWLK}, there is a separation of scales.
However, even in the 'tree' approximation the strength of the
generated field may be so high that its influence on the particle
("parton" \ ) trajectory should be taken into account. Presently available
transport codes, in particular the PHSD, allow one to consider
this effect and to get a consistent picture at the cost of an
essential simplification of the problem.

In the following we take the color charge as a unit vector and
specify the magnitude of the charge to be given by the coupling
constant $g$ [$\alpha_g = g^2/(4\pi)=0.3$], thus taking the
coupling strength out as an independent factor. This approximation
will be justified below in more detail.

The consideration of the non-Abelian charge as a function of time
leads to the generation of additional terms in the expressions for
the chromoelectric and chromomagnetic fields of color pointlike
charges which stem from the derivative of the color charge
vector with respect to the retarded time $t'$, i.e.,
\begin{eqnarray}
\label{24} &&\widetilde D=\frat{\partial \widetilde C} {\partial
t'}~, ~~~\frat{\partial \widetilde C}{\partial t}=\widetilde D
\frat{\partial t '}{\partial t}~, ~~{\vf\nabla} \widetilde C =
\widetilde D~{\vf\nabla} t '~, \nonumber \\ &&{\vf\nabla} \times
\widetilde C {\vf v} = {\vf\nabla} t '\times\widetilde D{\vf
v}+{\vf\nabla} t'\times\widetilde C \frat{\partial {\vf
v}}{\partial t '}~.\nonumber
\end{eqnarray}
The results for the non-Abelian chromo-fields are
\begin{eqnarray}
\label{25}
\widetilde {\vf E} &=& \frat{1} {4 \pi} ~
\left[\frat{\widetilde C} {R^2} \frat{(1 - \vf{v}^2) (\vf{n} -
\vf{v})} {(1 - \vf{ v} \vf{n})^3} + \frat{\widetilde C} {R}
\frat{\vf{n} \times (\vf{n} - \vf{v}) \times \dot{\vf{v}}} {(1 -
\vf{v} \vf{n})^3}  \nonumber \right. \\ &+& \left.
\frat{\widetilde D} {R} \frat{{\vf n} - {\vf v}} {(1-{\vf v}
{\vf n})^2} \right]_{t'}
\nonumber \\[- .2 cm]
\\[- .25 Cm]
 \widetilde {\vf H}&=&\frat{1}{4 \pi}~ \left[-\frat{\widetilde
C}{R^2} \frat{\vf{n} \times \vf{v}} {(1-\vf{v}\vf{n})^
3}(1-\vf{v}^2+\dot{\vf{v}}\vf{R})- \frat{\widetilde C}{R}\vf{n}
\times \dot{\vf{v}} \nonumber \right. \\ &-& \left.
\frat{\widetilde D}{R}\frat{{\vf n}\times {\vf v}}{(1-{\vf v}{\vf
n})^2} \right]_{t '}=\vf{n}\times \widetilde {\vf E}~. \nonumber
\end{eqnarray}
These additional terms look like radiation terms; however, one
cannot state that exactly because the change of the color charge
vector does not result from a particle displacement and such a
conclusion needs  further special analysis. It is difficult also
to compare directly these results with the Abelian case since the
chromoelectric and chromomagnetic field strengths are not
observable quantities. Accordingly, it is more consistent to
consider the energy density of chromoelectric and chromomagnetic
fields $\widetilde E^2/2$, $\widetilde H^2/2$  or the Wong force,
which is analogous to the Lorentz force in electrodynamics
\cite{Wong} and serves as a measure for the  momentum change of a
color particle in an external color field. Such a role could also
be played by a set of covariant functions, which define the
strengths of chromoelectric and chromomagnetic fields at the
different space-time points (in principle,  one could determine
the covariant gluon-parton distribution functions in the
Weizs\"acker-Williams ansatz).

In order to get well-settled approximate solutions  we need
further information that, e.g., can be obtained by solving the
problem of two slowly moving color charges~\cite{Khrip,GMS}.
Assuming one particle to carry a color charge $\widetilde P$ and
the other particle  a charge $\widetilde Q$ we are allowed to
simplify the consistency equations (\ref{13}) for the pointlike
sources to the following form~:
\begin{eqnarray}
\label{15}
&& \dot {\widetilde P} = g~\widetilde \varphi (t,\vf{x}_1) \times
\widetilde
P~,\nonumber\\
[- .2 Cm] \\[- .25 cm]
&& \dot {\widetilde Q} = g~\widetilde \varphi (t,\vf{x}_2)
\times\widetilde
Q~,\nonumber
\end{eqnarray}
where the dot denotes the time derivative, $\vf{x}_1$ and $\vf{x}_2$
are the positions of the pointlike charges at time $t$, and
$\widetilde\varphi(t,\vf{x}_i)$ is the value of the scalar potential
at the charge location. An immediate consequence of
Eqs.~(\ref{15}) is the conservation of the modulus of the particle
color charge $\dot{\widetilde P^2}=0$ and $\dot{\widetilde
Q^2}=0$. An analysis of the perturbation series in the coupling
constant $g$ and in particle velocity $v/c$ shows that if the scalar
potential $\widetilde \varphi $ is spanned by the two color vectors
defined above
\begin{equation}
\label{16} \widetilde \varphi = \varphi_1~\widetilde P +
\varphi_2~\widetilde Q~,
\end{equation}
(in particular, the first term of the iteration series looks
similarly), then the vector potential $\widetilde {\vf A}$
(generated by the charges) is spanned by the vector product of charges
only
\begin{equation}
\label{17} \widetilde {\vf A}={\vf a}~\widetilde P \times\widetilde Q~.
\end{equation}
In fact, Eqs.~(\ref{16}) and (\ref{17}) fix the Coulomb gauge.
Moreover, as a consequence of the vector algebra rules,
contributions to the scalar potential are given only through the
terms spanned by two charge vectors $\widetilde P$, $\widetilde
Q$. Finally, the equation is factorized and we have the following
 system of equations for the  potential components
$\varphi_1$, $ \varphi_2 $ and the vector field ${\vf a}$
\begin{eqnarray}
\label{18}
&& {\vf D}{\vf D}~\Phi =\delta~, \nonumber \\[-.2 cm]
\\[-.25 cm]
&&\nabla \times \nabla \times {\vf a}-g~{\vf j}=0~, ~~~{\vf
j}=\Phi J{\vf D}\Phi~, \nonumber
\end{eqnarray}
where the column $ \Phi = \varphi-\! \stackrel{*} {\varphi}$ with
$\varphi^{\mbox {\tiny T}}=\|\varphi_1, \varphi_2\|$,
${\stackrel{*}{\varphi}}^{\mbox {\tiny
T}}=\|\stackrel{*}{\varphi}_1, \stackrel{*}{\varphi_2}\|$,
$\stackrel{*}{\varphi_1}=\varphi_1({\vf x}_2)$, and
$\stackrel{*}{\varphi_2}=\varphi_2({\vf x}_1)$ is the potential at
the points of the charge location, $\delta^{\mbox {\tiny T}} =
\|\delta({\vf x}-{\vf x}_1), \delta({\vf x}-{\vf x}_2)\|$, \ ${\vf
D}_{kl}=\nabla \delta_{kl}+g{\vf a}C_{kl}$, \\$k,l = 1,2$ is the
covariant derivative and $C$, $J$ are the following matrices~:
\begin{figure*} 
\begin{center}
\includegraphics[width=0.45\textwidth,clip=true] {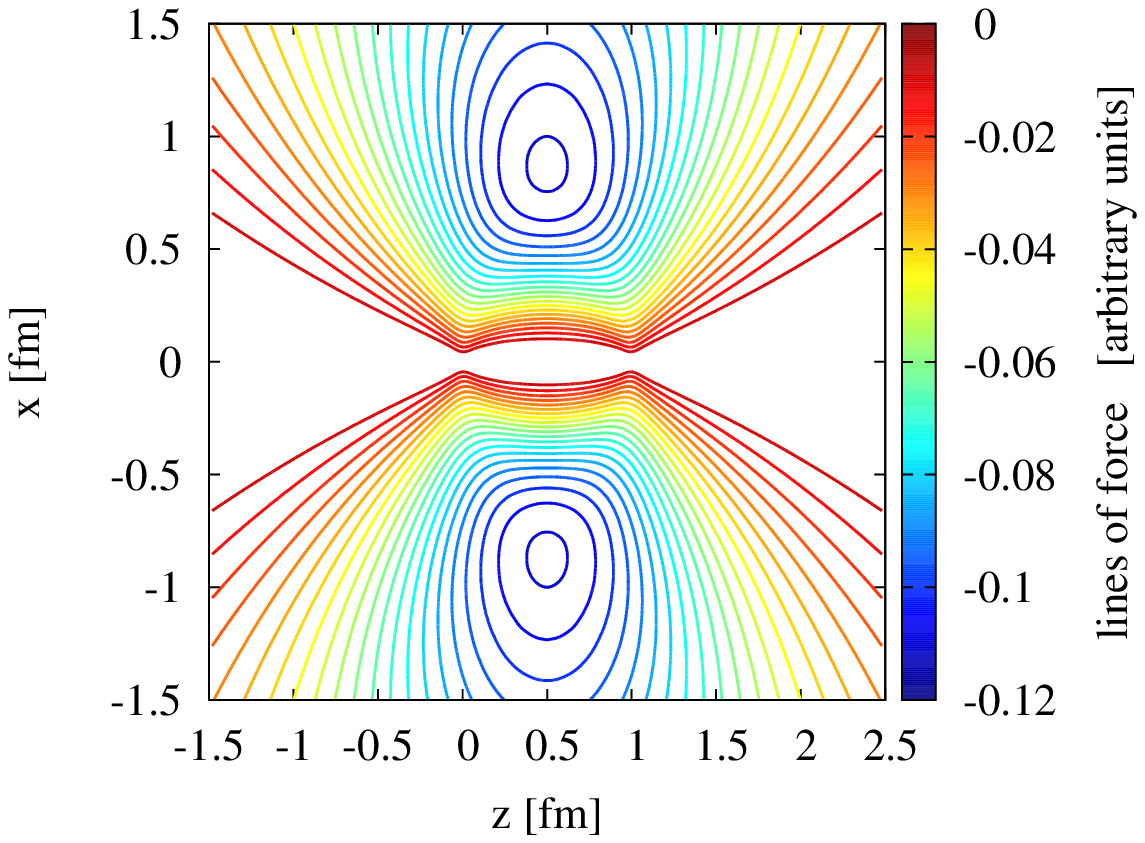}
\hspace{0.7cm}
\includegraphics[width =0.45\textwidth,clip = true]{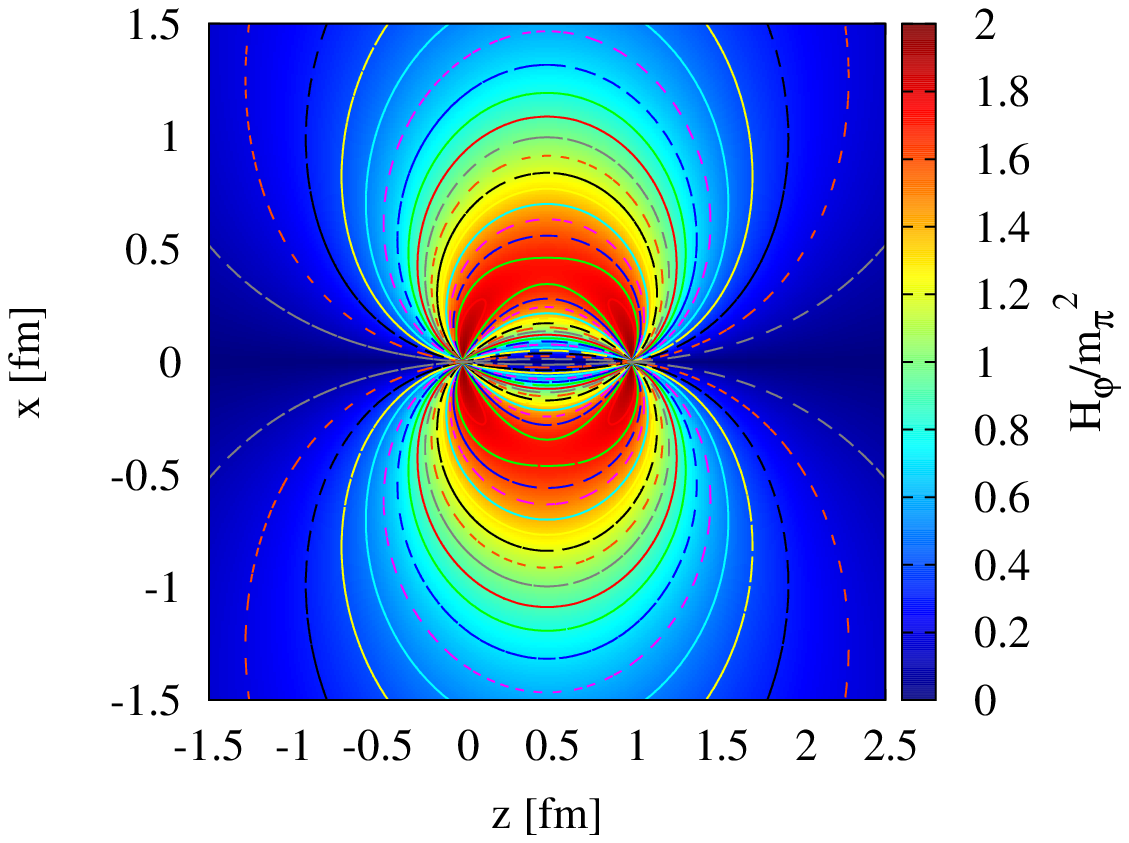}
\vspace{-1mm}
\begin{picture}(0,0)(0,0)
 \put(-338,135){\small{({\bf a})}}
 \put(-83,135){\small{({\bf b})}}
\end{picture}
\caption{(Color online) Schematic view of the lines
of force of the vector field ${\vf a}$ (a) and the
isolines of the chromomagnetic field component $H_\varphi=rot \
{\vf a}$ (b). The results are given for the cylindrical
symmetry with two charges located at the points with z=0 and 1.}
 \label{fad}
\end{center}
\end{figure*}

\vspace{0.25cm}
\begin{center} \label{rot}
\parbox[b] {3.in} {$
C = \left\| \begin{array} {rr} -(\widetilde P \widetilde Q) &
                            - (\widetilde Q \widetilde Q) \\
 (\widetilde P \widetilde P) &
                            (\widetilde P \widetilde Q)
\end{array}
\right\|, $ ~~~$ J = \left\| \begin{array}{rr}
 0 & 1 \\
-1 & 0
\end{array} \right\|. $}
\end{center}
\vspace{0.25cm} \noindent Here the parentheses denote the scalar
product of charge vectors and $\delta$ is the $\delta$-function
source of unit intensity. Then Eqs.~(\ref{15}) achieve  the
following form~:
\begin{eqnarray}
\label{19} && \dot {\widetilde P} = g~\stackrel{*}{\varphi_2}~
\widetilde Q \times \widetilde P~, \nonumber \\[- .2 cm]
\\[- .25 cm]
&& \dot {\widetilde Q} = g~\stackrel{*}{\varphi_1}~
\widetilde P \times \widetilde Q~. \nonumber
\end{eqnarray}
This system of Eqs.~(\ref{19}) describes the rotation of color
charges with respect to  the constant vector $\widetilde \Omega =
\stackrel{*}{\varphi}_1\widetilde
P+\stackrel{*}{\varphi}_2\widetilde Q$ with the frequency
$g|\widetilde \Omega|$. The matrix $C$ is conserved $d C/dt=0$, so
the potentials of the scalar $\varphi$ and vector field ${\vf a}$
are quasi-static, indeed.

The set of Eqs.~(\ref{18}) has a transparent physical
interpretation. The color field, generated by two pointlike
sources, is the source of a color charge itself. Thus, a
self-consistent environment of color charges and the corresponding
currents is established in the space around the charges. The
solutions of this system were investigated in detail both
analytically and numerically in Ref.~\cite{GMS}. There, it was
demonstrated that the interaction between color charges is
Coulomb-like. If the coupling constant is not large, i.e., ${g^2}/(4
\pi) <\sqrt{2}$, the corresponding solution can be well
approximated by the Coulomb potentials $\varphi_{1,2}=1/(4 \pi
|{\vf x}-{\vf x}_{1,2}|)$ with the vector field ${\vf a}$ that is
generated by these potentials in the next iteration in $g$. This
is illustrated in Fig.~\ref{fad} by drawing the "lines of constant
force" \ for $\vf{a}$ [see Fig.~\ref{fad}(a)] and equi-potentials
 for the induced chromomagnetic field $H_\varphi=rot \
{\vf a}$ [Fig.~\ref{fad}(b)], respectively. A line of force is a directed
curve in the color field such that a forward tangent at any point
shows the direction of the chromofield intensity. It is seen that
the vector field of two color charges looks like the field of a
permanent magnet with poles being placed at the points ${\vf
x}_{1},{\vf x}_{2}$. The vector field ${\vf a}$ develops a
constant longitudinal component $|{\vf a}_{\|}|=1/(4\pi|{\vf
x}_1-{\vf x}_{2}|)$ only along the straight line connecting the
sources at $z_{1}=0$ and $z_{2}=1$, as demonstrated in
Fig.~\ref{fad}, where the current lines of the vector field ${\vf
a}$ are presented. The field drops sharply perpendicular to the
line on which the color charges are located. The chromomagnetic
field includes a single (vortex) component and, for example, for
color charges at the points $0$ and $1$ on the  axis $z$, is given
by
 \begin{eqnarray}
\label{Hphy} \widetilde H_\varphi &=& \frat{g^3}{(4
\pi)^2}~\frat{1}{r}~
\left(\frat{1-\mu}{1+\mu}\right)^{1/2}~\left(1-\frat{1+r} {(1-2r
\mu + r^2)^{1/2}}\right)\nonumber \\ & \cdot& \widetilde P \times
\widetilde Q \equiv H_\varphi \ \widetilde P \times \widetilde Q~,
\end{eqnarray}
where $r$, $\mu$ are the spherical coordinates measured from the
origin and $\mu =\cos\theta$ with the angle $\theta$ measured
with respect to the $z$ axis.  The contour lines of the $H_\varphi$
component of the chromomagnetic field are presented in Fig.~\ref{fad}(b).
It is noteworthy that the lines of
constant force and equi-potentials seem to be mutually orthogonal
at each $z,r$ point. Equation~(\ref{Hphy}) becomes more
complicated for other choices of the charge location.

The total energy concentrated in the color field
may be estimated as
\begin{eqnarray}
\label{20} {\mathrm E}&=&\int d^3x~ \frat{\widetilde {\vf
E}^2+\widetilde {\vf H}^2}{2} \nonumber \\ &\simeq&
\alpha_g~\frat{(\widetilde P\widetilde Q)}{|{\vf  x}_1-{\vf
x}_{2}|}+ \alpha_g^3~I~\frat{ (\widetilde P\times\widetilde
Q)^2}{|{\vf  x}_1-{\vf x}_{2}|}~,
\end{eqnarray}
with $\alpha_g=g^2/(4 \pi)$, $I=(6-\pi^2/2)/4$; the terms of the
source self-interaction are not included here. These estimates of
the energy density  of the chromomagnetic and chromoelectric
fields clearly demonstrate the relative contributions of the first
and subsequent iterations as well as the dominant role of the
first iteration in the coupling constant relevant for actual
applications in ultrarelativistic nucleus-nucleus collisions.

A closer analysis shows that the potentials $ \varphi_i $ have a
singularity in the locations of the charges. The limiting values
$\stackrel{*}{\varphi}_i$
(cf. Eqs.(\ref{19})) depend on the path by which one arrives at
the charge position. For example, for one of the charges in the
origin we get
\begin{eqnarray}
&&\varphi_1 =-\frat{1}{4 \pi}~\frat{1}{|{\vf x}|}-
\frat{g^3}{4 \pi}~a\mu~(\widetilde P \widetilde Q)+\dots \nonumber \\
&&\varphi_2 =-\frat{1}{4 \pi}~\frat{1}{|{\vf x}-{\vf x}_2 |}+
\frat{g^3}{4 \pi}~a\mu~(\widetilde P \widetilde P)
+ \dots \nonumber
\end{eqnarray}
where $a$ is the value of the longitudinal component of the
vector field ${\vf a}_\|$ at the
location point of the charge. Formally, the presence of
singularities leads to a contradiction with the initial
assumptions of deriving Eqs.~(\ref{19}) because now a mismatch of
the color vector rotation on the charge itself is possible. In
fact, it signals the necessity of a more accurate fixing of
boundary conditions on the charge (this problem is akin to similar
problems of diffraction theory). In particular, if these
conditions are formulated in such a way that the vector field
${\vf a} $ does not penetrate into the sources, one can avoid
these difficulties~\cite{mm}. However, the approximate solutions
are still quite appropriate to be used for the coupling constants,
which are  of interest for applications. Then it can be shown that
the solutions of Eq.~(\ref{18}) with a boundary condition that
leads to expelling the vector field from the charge should be
qualitatively restructured when the coupling constant reaches the
threshold values
\begin{equation}
\label{o1}
\Pi =\frat{g^2}{(4 \pi)^2}~\widetilde P^2\leq l(l +1)~,
\end{equation}
where $l = 0,1,2,\dots $ are integers (of angular momentum) and
$l=0$ corresponds to the situation with the penetrating vector
field. In addition to these critical values there is another one:
it results from the same characteristic equation that we have used
to obtain the limit (\ref{o1}) and is associated with the
singularity power of (\ref{18}) solutions in the  neighborhood of
the origin, where the source of the color charge $ \widetilde P $
is located. Searching for the  solution  of
$(\varphi_i-\stackrel{*}{\varphi}_i)$ and the vector field component
$a$ in the form of $r^\sigma$ we obtain for the exponent $\sigma$
$$ \sigma = -\frat{1}{2}\pm \frat{\left((2l+1)^2-4 \pi \right)^{1/2}}{2}~.
$$
The condition to treat only the real-valued solutions
\begin{equation}
\label{o2} \Pi \leq (l+1/2)^2
\end{equation}
leads to the constraint widely discussed some time ago \cite{mnd}.
It was asserted that when reaching this threshold the non-Abelian
fields develop  instability and, as a consequence, the color charge
becomes completely screened. However, one can notice that the
critical values (\ref{o1}) are always reached before the limit
(\ref{o2}) becomes effective and, therefore, a rearrangement of the vector
field takes place before the complete color charge screening
develops. The proper time-consuming calculations are possible only
by applying numerical schemes with the vanishing divergence.
Fortunately, such effects occur for coupling constants which
apparently are of little interest for the applications.

The scalar component of Eqs.~(\ref{18}) can be written in the form
$${\vf\nabla}{\vf E} =\delta-{\vf a}C{\vf E}~,$$
where  a column of two components of the chromoelectric field
${\vf E}={\vf D}\Phi$ has been introduced. The last term includes
the vector field ${\vf a}$ and can be interpreted as a charge
density cloud (column of charges $G_1 $, $ G_2$) of the
gluon field $$ G =-{\vf a}C{\vf E}~.$$
Intuitively, it appears plausible that if the color sources are
antiparallel, then the charge of the non-Abelian field could be
zero. Indeed, if $ \widetilde P =-\widetilde Q $, then by
calculating the charge density components $G_1=G_2={\vf
a}{\vf\nabla}(\Phi_2-\Phi_1)$, it becomes clear that the charge of
the non-Abelian field indeed vanishes $\widetilde G=G_1\widetilde
P+G_2 \widetilde Q=\widetilde 0$. In contrast, we expect that the
non-Abelian field charge becomes maximal if the color charges of
the particles are parallel $ \widetilde P = \widetilde Q $. In
this situation $G_1=- G_2={\vf a}{\vf\nabla}(\Phi_1+\Phi_2)$,
where, contrary to expectations, the non-Abelian field charge gets
zero $\widetilde G=\widetilde 0$ again. Note, however, that the
charge of the non-Abelian field will be nontrivial for other
configurations.

It is important to notice that the energy of the system of the
non-Abelian field and the particles can be written (if based on the
solutions of Eqs.~(\ref{18})) in the following form:
\begin{eqnarray}
\label{o3}  {\mathrm E}&=&\int d^3 x~\frat{\widetilde {\vf
E}^2+\widetilde {\vf H}^2}{2}\nonumber \\ &=& \frat{1}{2}~\int d^3
x~\left(-\widetilde \varphi~\widetilde \delta+ \widetilde
\varphi~\widetilde G\right)~,
\end{eqnarray}
where in accordance with Eq.~(\ref{16}) $\widetilde\delta=\delta_1
\widetilde P+\delta_2\widetilde Q$, i.e., the chromomagnetic
component effectively transforms into a charge. This form
reminds us of the well-known expression of electrodynamics for the
electrostatic energy of a system of  charges
$${\mathrm E}=-\frat{1}{2}\sum_i \varphi_i e_i~, $$
where $\varphi_i$ is the potential  (that can already be
considered as an observable quantity) at the location point of the
charge $e_i$. Comparing these expressions we are able to conclude
clearly about the role of the non-Abelian charge cloud and, taking
into account the sign of its contribution to the energy (see
Eq.~(\ref{20}), for example) we find that   extra repulsion is
developed in the system of non-Abelian charges. We speculate that
this extra repulsion might also lead to enhancement of the
elliptic flow $v_2$ of hadrons in relativistic heavy-ion
collisions which has not been considered so far.

In view of the classical field dynamics we realize that
Eq.~(\ref{o3}) serves as the only source of information on the
forces acting in the system. It is interesting to remark that, e.g.,
the Wong force (invented to describe the dynamics of
color charges in an external field) is associated only with the
first term $\sim \widetilde \varphi \widetilde \delta$ of
Eq.~(\ref{o3}), as it should  be in the chromostatic case, since
it is the only combination permitted to construct  properly its
covariant momentum extension. It can  also be constructed by counting the
Wong force for the first and second charge explicitly (using the
definition of the chromoelectric field strength $\widetilde {\vf
E}$) :
\begin{eqnarray}
&& {\vf F}_P = \widetilde P \widetilde {\vf E}({\vf x}_1)=
{\vf\nabla} \varphi_1 |_{{\vf x}_1}~\widetilde P^2 +
{\vf\nabla} \varphi_2 |_{{\vf x}_1}~(\widetilde P \widetilde Q)~,
\nonumber \\
&& {\vf F}_Q = \widetilde Q \widetilde {\vf E}({\vf x} _2) =
{\vf\nabla} \varphi_1 |_{{\vf x}_2}~ (\widetilde P \widetilde Q) +
{\vf\nabla} \varphi_2 |_{{\vf x}_2}~ \widetilde Q^2~. \nonumber
\end{eqnarray}
Singular contributions describe  self-interactions and should
be regularized as in electrodynamics. Equation~(\ref{o3})
demonstrates explicitly that  the superposition principle for a
color charge is  non-applicable in the general situation and one
has to accurately account for the non-Abelian field cloud energy.
The structure of the cloud environment and, hence, the system
energy depends, generally speaking, on the choice of the boundary
condition for the charge.

It is relevant to notice that the iterative perturbation series can
be analyzed [when dealing with the Yang-Mills systems (\ref{12})]
even when discarding  the compatibility condition (\ref{13}). The
correct construction of the iterative series inevitably leads to the
conclusion about the nonconservation of classical color charge. This
property is essentially different from naively expected small
corrections to the Abelian solutions because of the nonlinearity of
the Yang-Mills equations. In a sense this conclusion can be used in
the inverse manner that the solutions are automatically
correct with accuracy of an order of $g$ (in reality $g^3$, see
(\ref{o3})) if the compatibility conditions are taken into account.
In the non-Abelian case it is technically difficult to operate with
the different gauges and besides, the set of physically justified
(gauge invariant) observables is less than in electrodynamics.

Since the  strengths of chromoelectric and chromomagnetic fields
are not  proper observables,  the problem  arises in interpreting
their solutions. Actually, in the example discussed above the
gauge is implicitly fixed by selecting a special type of
solutions. If one makes a choice in favor of the rest frame of the
rotating charges, the potentials $\varphi_{1,2}$ approach some
constants at  spatial infinity. Such a gauge is inconvenient to be
used in  numerical calculations and our experience suggests that
the  Coulomb solutions are most informative.

Adding even one more color charge makes it impossible to
investigate the problem at the  same level of consistency as for
the two-particle problem because the compatibility equations
(\ref{13}) cannot be integrated anymore. Nevertheless, it seems
reasonable to consider that the system energy is given by the same
expression (\ref{o3}) with the Coulomb interaction of the
components. The charge contributions of non-Abelian fields
(diffusive color clouds) are small  for the coupling constants of
actual interest because in the relativistic heavy-ion situation it
is impossible to factorize the equations in a general form and the
 problem  arises of how to construct an approximate solution.

\section{The field of two color charges}
Let us first consider the field of two color charges $\widetilde P$,
$\widetilde Q$ moving along the $z$-axis towards each other  with
velocities $v$ and $-w$, respectively. If their encounter is
supposed to take place at time $t=0$, their coordinates in the
laboratory system are given as $z_1=vt, z_2=-wt$. Obviously, when
the particles are far from each other their interaction can be
assumed weak and their charges remain constant with high accuracy.

\begin{figure}
\includegraphics[width = 7.cm, clip = true] {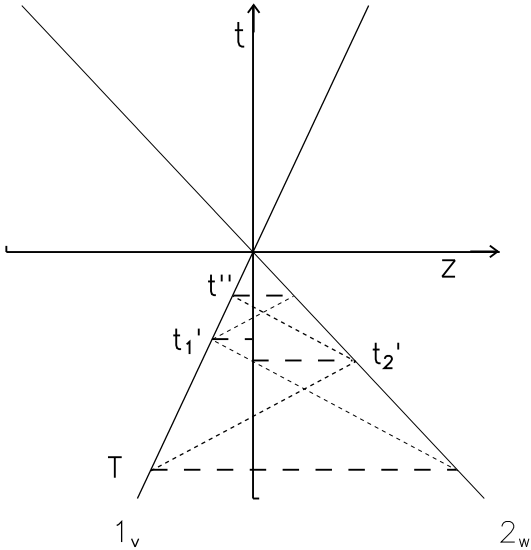}
\caption{The time scheme for the meeting of two color particles.
The solid lines are the particle trajectories, the dashed lines
show projections on the time axis, the dotted lines are the light
signals. Here $T$ is the initial time of the interaction when the
particles become closer than the distance $D=1$ fm; $t_1'$ is the
arrival time of the signal to the first particle about the
presence of the color charge $\widetilde Q_T$ at $1$ fm; similarly
for the second particle; $t_2'$ is the time when
 the charge $\widetilde P_T$ is seen  at the same distance, where
$\widetilde P_T $, $ \widetilde Q_T $ are the charges before
entering the interaction zone, i.e., before the time $T$. Until the
time $t''$, the first and the second particles rotate with respect to
constant color vectors $ \widetilde Q_T$, $\widetilde P_T$.  }
 \label{f1}
\end{figure}

Adapting the typical scale of non-Abelian (strong) interactions of
about $1$ fm we assume  that the interaction is switched on just
at this distance (scale) denoted by $D$. Obviously, this generates
the characteristic time scale in the problem. Then the first
"milestone" appears when the particles enter the interaction area,
i.e.,
\begin{equation}
\label{26} T =-\frat{D}{v+w}~.
\end{equation}
The second time instant of importance is $t_1'$ when a signal of the
appearance of charge $\widetilde Q_T$ at the distance of $1$ fm
reaches the first particle (with the color charge $\widetilde P$).
Similarly for the second particle, the time  when the charge
$\widetilde P_T$ comes at the same distance is denoted by $t_2'$,
where $\widetilde P_T$, $\widetilde Q_T$ are the color charges
before entering the zone of interaction, i.e., before the time $T$,
\begin{equation}
\label{27}
t_1'=\frat{1-w}{1+v}~T,~~t_2'=\frat{1-v}{1+w}~T.
\end{equation}
These expressions are easily extracted from the scheme in
Fig.~\ref{f1} by writing out the relations for the arrival of the
"light" \ signals and charges at the points of interest. In the
figure, these events take the form of the corresponding triangles.
Just at these times the color charges start to rotate with respect
to their constant color vector, which is the vector peculiar to the
partner charge before it entered the interaction area. This regime
is going on up to the time
\begin{equation}
\label{28}
t''=\frat{1-w}{1+v}\frat{1-v}{1+w}~T~,
\end{equation}
which is the same for both particles. It is also the time
necessary for a signal to reach the partner providing information  on
the beginning
of rotation [see Eqs.~(\ref{19})] starting from its asymptotic
charge value $\widetilde P_T$, ($\widetilde Q_T$). It is
noteworthy that the same approach is used for formulating properly
the Cauchy problem for retarded equations~\cite{du}.

The velocity of a relativistic particle is determined by the relation
\begin{equation}
\label{29} v=\left(1-\frat{m^2}{{\cal E}^2}\right)^{1/2} \simeq
1-\frat{m^2}{2~{\cal E}^2}~,
\end{equation}
where $m$ is the particle mass and $ {\cal E} $  its  energy. It
allows us to estimate the order of magnitude for the
characteristic time in the problem as
$$ t'\sim \frat{m^2}{{\cal E}^2}~T,~~t'' \sim \frat{m^4}{{\cal
E}^4}~T~.$$
The collision energies of present heavy-ions facilities (RHIC and
LHC) allow us to estimate the corresponding factors as $ {\cal E} /
m \sim 10 - 10^{2}$ and more.
 In Secs V and VI, where charge-dipole and dipole-dipole
scattering is considered, another interesting scale arises. It is
the distance $\delta$     between the dipole charges  and we
assume it to be of the order of the interquark distance in the
nucleon, i.e.  $\delta=1$ fm. Then in the laboratory system
the dipole size (due to  Lorentz contraction) will be
$$\Delta=\delta~(1-v^2)^{1/2} \sim \delta~\frat{m}{{\cal E}}~.$$
Let us denote the time corresponding to this scale by $t_3 \sim
m/{\cal E}$ (see below). Thus, for the relativistic problem of
interest we obtain the following temporal hierarchy of interaction
stages
$$t''\ll t '\ll t_3 \ll T.$$

Due to the chosen geometry of the problem an approximate solution
of the Yang-Mills equations for two color charges can be
represented as the following superposition [see also (\ref{14})]~:
\begin{equation}
\label{30}
\widetilde \varphi=[\varphi_1 \widetilde P]_{t'}+[\varphi_2 \widetilde
Q]_{t'}~,
~~\widetilde A_z=v~[\varphi_1 \widetilde P]_{t'}-w~[\varphi_2 \widetilde
Q]_{t'}~.
\end{equation}
The scalar potentials $\varphi_1$ and $\varphi_2$ can be
specified as:
\begin{equation}
\label{31}
\varphi_1=\frat{1}{4 \pi}~\frat{g}{R_1-v~R_1}~,~~
\varphi_2=\frat{1}{4 \pi}~\frat{g}{R_2-w~R_2}~.
\end{equation}
\begin{figure}
\includegraphics[width = 7.cm, clip = true] {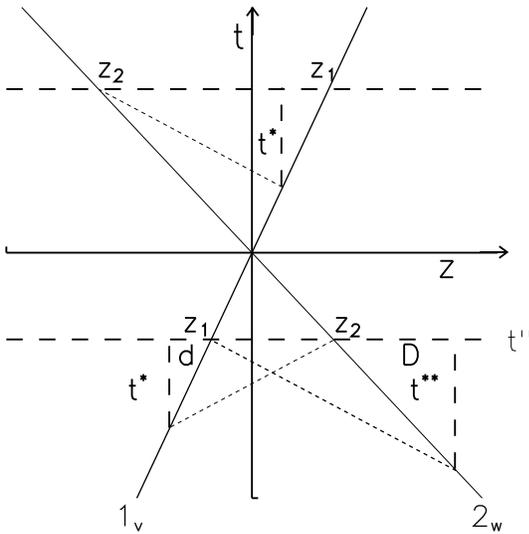}
\caption{The scheme for determining the retardation time for the
collision of two charges. Here $z_1$ and $z_2$ are the current
positions of the particles at two certain time moments. The charge
of the first particle is characterized by the retarded time $
t^{**} $ as well as by a segment $D$ which measures the distance
between the particles for this  time.  The parameters $ t^{*} $
and $d$ for the second particle  have a similar meaning (see text). } \label{f2}
\end{figure}
The factor $g$ is taken out since the normalization of the color
charge vectors of the particles is  chosen as unity. The
consistency conditions (\ref{23}) include the potentials at the
location point of the particles. Let $ z_1 $ and $ z_2 $ be the
coordinates of the particles at some moment (see Fig. \ref{f2}).
From this picture one can conclude that the potential values of
interest are determined by the distance
$$R_1=d+z_2-z_1=t^*~,~~R_2=D+z_2-z_1=t^{**}~ $$
with $d=vt^*~$ and $D=wt^{**}$. Using these relations, we obtain
$$\varphi_1=\frat{1}{4 \pi}~\frat{g}{z_2-z_1}~,
~~\varphi_2=\frat{1}{4 \pi}~\frat{g}{z_2-z_1}~.$$ Thus, the
Lorentz factors are compensated in such a way that the scalar
potentials at the charge location  can be simply expressed in
terms of the distance between the particles in the current moment.
Also, taking into account the form of the vector potential, we
arrive at the following expression for the compatibility
conditions~:
\begin{eqnarray}
\label{32}
&&\dot{\widetilde P}=\alpha_g~\frat{1+v~w}{|z_1-z_2|}~
\widetilde Q (t-t^{**}_{12}) \times \widetilde P~, \nonumber \\[- .2 cm]
\\[- .25 Cm]
&&\dot{\widetilde Q}=\alpha_g~\frat{1+v~w}{|z_1-z_2|}
~\widetilde P(t-t^*_{21}) \times \widetilde Q ~. \nonumber
\end{eqnarray}
The retarded time can be obtained (using Fig.~\ref{f2}, for example), as
\\
$$t^*_{21}=-\frat{v+w}{1-v}~t~,~~t^{**}_ {12}=-\frat{v+w}{1 -w}~t~,~~
{\mbox{for}}~t<0~.$$
 For $t>0 $ (in the above formulas) the following replacement should be
made: $ v, w \to -v, -w $. In the electromagnetic case, the charge
is conserved and its retarded time is defined simply as
$t=t'+R(t')/c$~\cite{Ton}. Below we will apply the reduced notation
for the retarded time without any further explanation.

It is convenient to introduce  an auxiliary variable $\chi=-\ln|t|$,
by means of which Eqs.~(\ref{32}) are reduced to
\begin{eqnarray}
\label{33} &&\widetilde P'=\omega~\widetilde
Q(\chi-\Delta^{*})\times\widetilde P~,~~~( t<0) \nonumber
\\[-.2cm]
\\[-.25cm]
&&\widetilde Q'=\omega~\widetilde P(\chi-\Delta^{**})\times\widetilde
Q~,\nonumber
\end{eqnarray}
for negative times, and
\begin{eqnarray}
\label{34}
&&\widetilde P'=-\omega~\widetilde Q(\chi+\Delta^{**})\times\widetilde
P~,~~~(t>0)
\nonumber \\[-.2 cm]
\\[-.25cm]
&&\widetilde Q'=-\omega~\widetilde P(\chi+\Delta^{*})
\times \widetilde Q~, \nonumber
\end{eqnarray}
for positive times. Here the prime denotes differentiation with
respect to $\chi$,
$$\omega=\alpha_g~\frat{1+v~w}{v+w}~,~~
\Delta^*=\ln \frat{1+v}{1-w}~,~~\Delta^{**}=\ln \frat{1+w}{1-v}~. $$

\begin{figure*}
\begin{center}
\includegraphics[width=0.45\textwidth, clip = true] {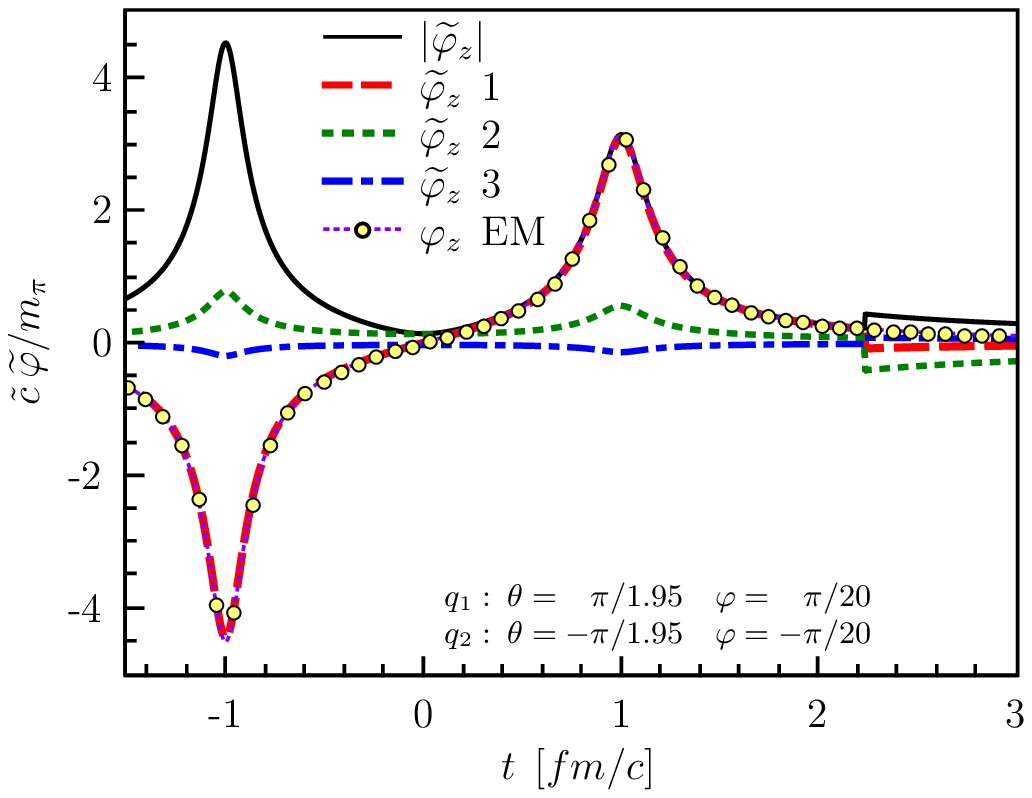}
\hspace{0.5cm}
\includegraphics[width =0.45\textwidth, clip = true] {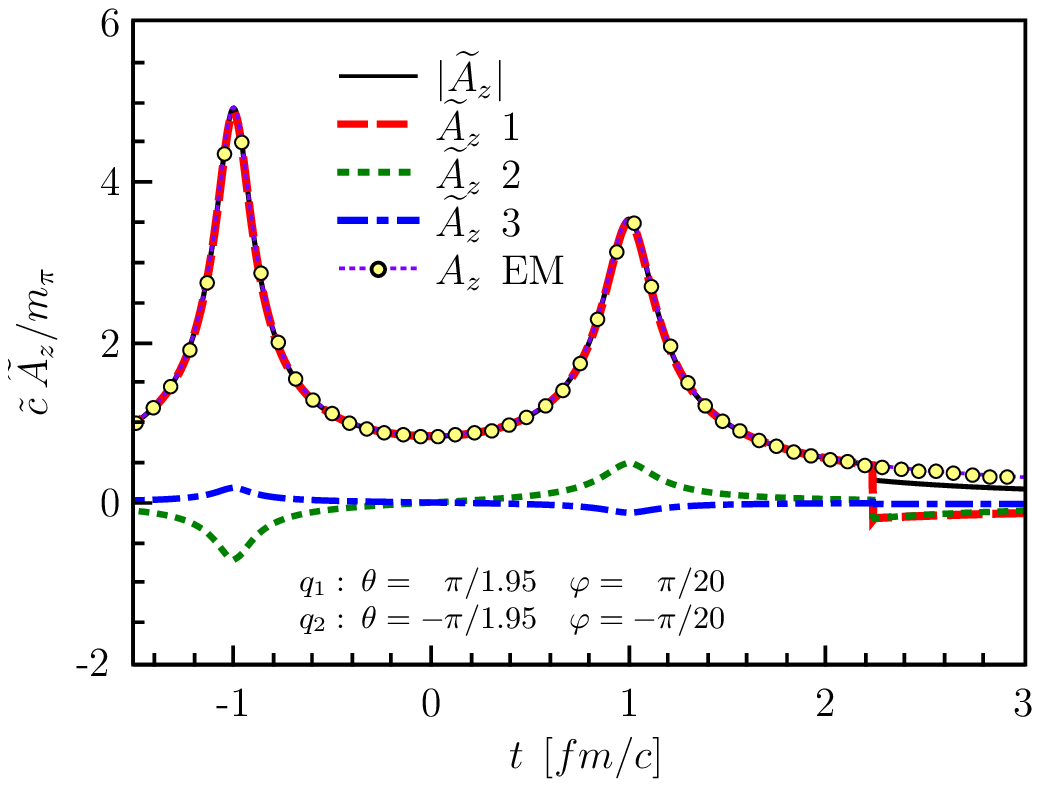}
\vspace{-3mm}
\begin{picture}(0,0)(0,0)
\put(-285,150){\small({\bf a})}
\put(-30,150){\small({\bf b})}
\end{picture}
\caption{(Color online) Time evolution of three
color components for the scalar $ \widetilde \phi $ (a)
and $z$ component of the vector potential $ \widetilde A_z $
(b)  for two oppositely directed charges in color space
as a function of time is given by the dashed lines. The solid line
shows the absolute value of the isovector potential $ | \widetilde
\phi |$. The electrodynamic potential with charges $ \pm e$
corresponding to the coupling constant $ \alpha_e = 0.3 $ is
displayed by open circles.} \label{f3}
\end{center}
\end{figure*}
\begin{figure*}
\begin{center}
\includegraphics[width =0.45\textwidth, clip = true] {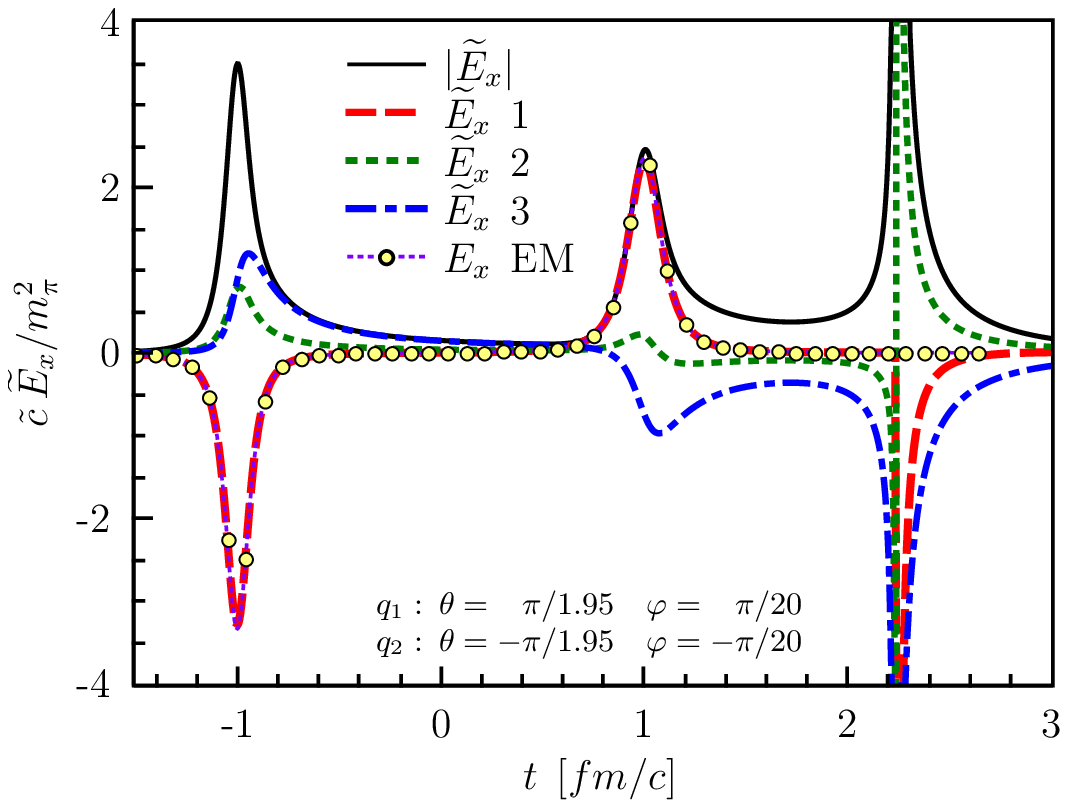}
\hspace{0.5cm}
\includegraphics[width =0.45\textwidth, clip = true] {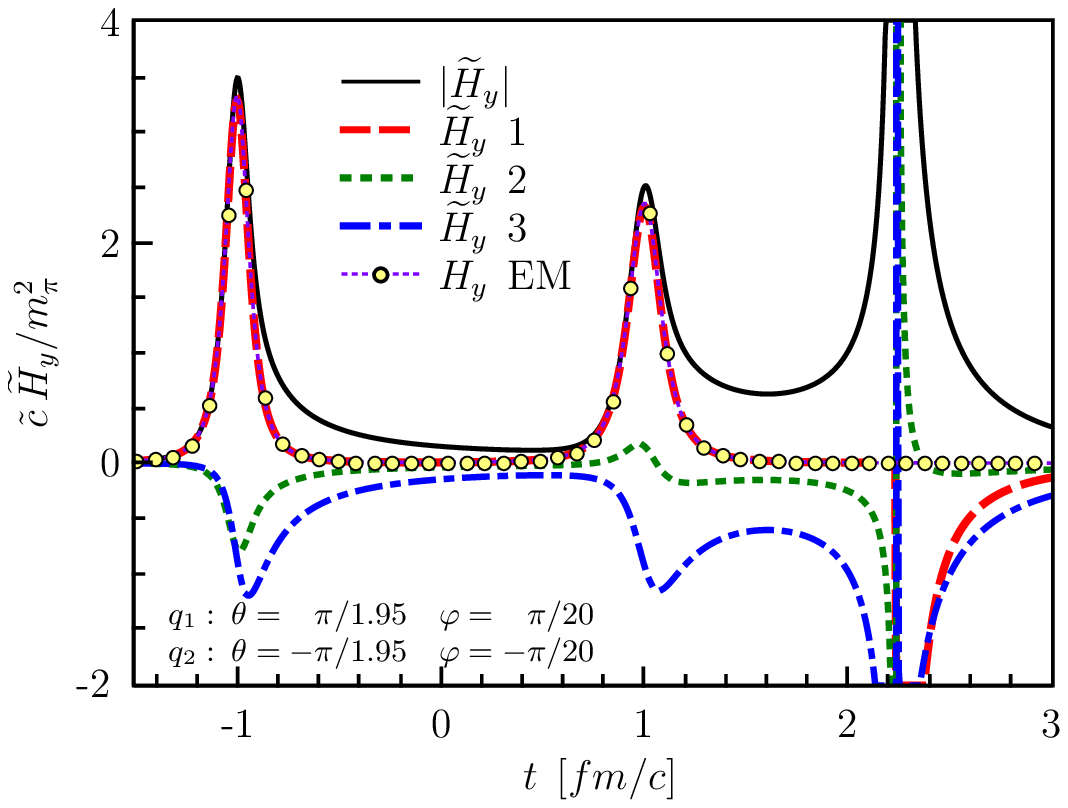}
\vspace{-3mm}
\begin{picture}(0,0)(0,0)
\put(-285,150){\small({\bf a})}
\put(-30,150){\small({\bf b})}
\end{picture}
\caption{(Color online) Three color components of
the chromoelectric $\widetilde E$ (a)  and chromomagnetic
$\widetilde H $ field strength (b) for two moving color
charges of  opposite signs as a function of time are given by
the dotted lines. The solid line shows the absolute value of the
isovector potential $|\widetilde E|$. The open circles  show the
electrodynamic vector field corresponding  to charges $ \pm e $
and coupling constant $\alpha_e=0.3$. }
 \label{f4}
\end{center}
\end{figure*}
\begin{figure*}
\begin{center}
\includegraphics[width =0.45\textwidth, clip = true] {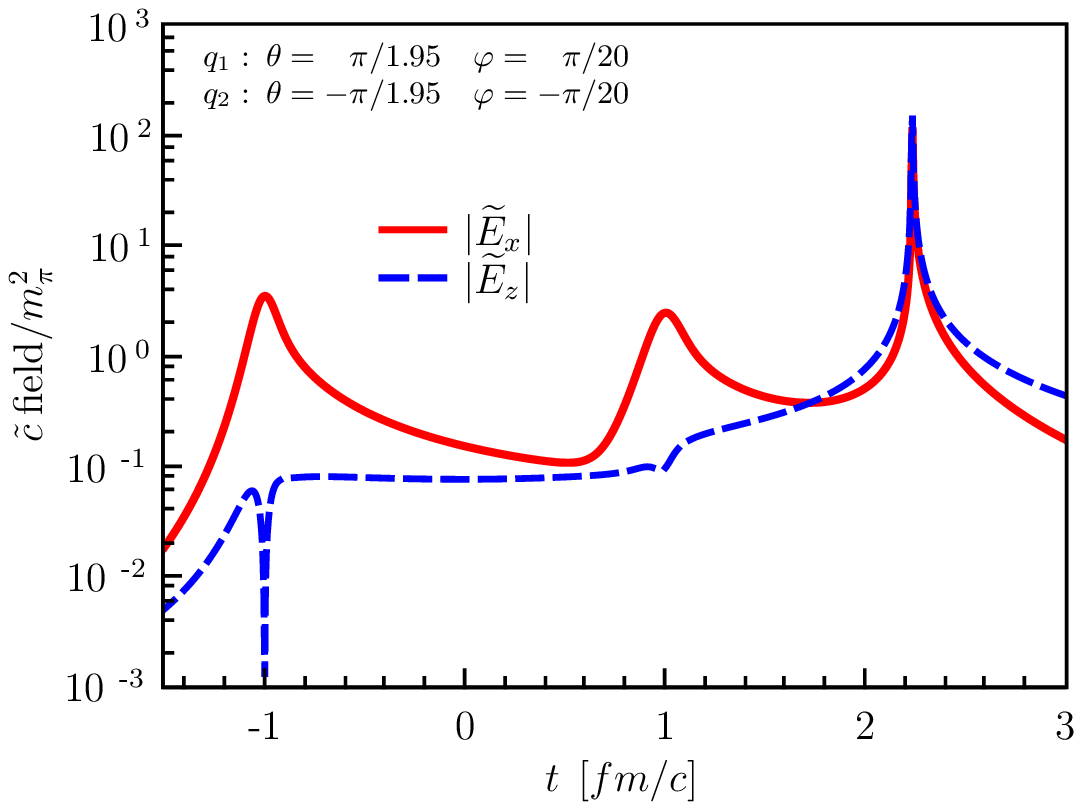}
\includegraphics[width =0.45\textwidth, clip = true] {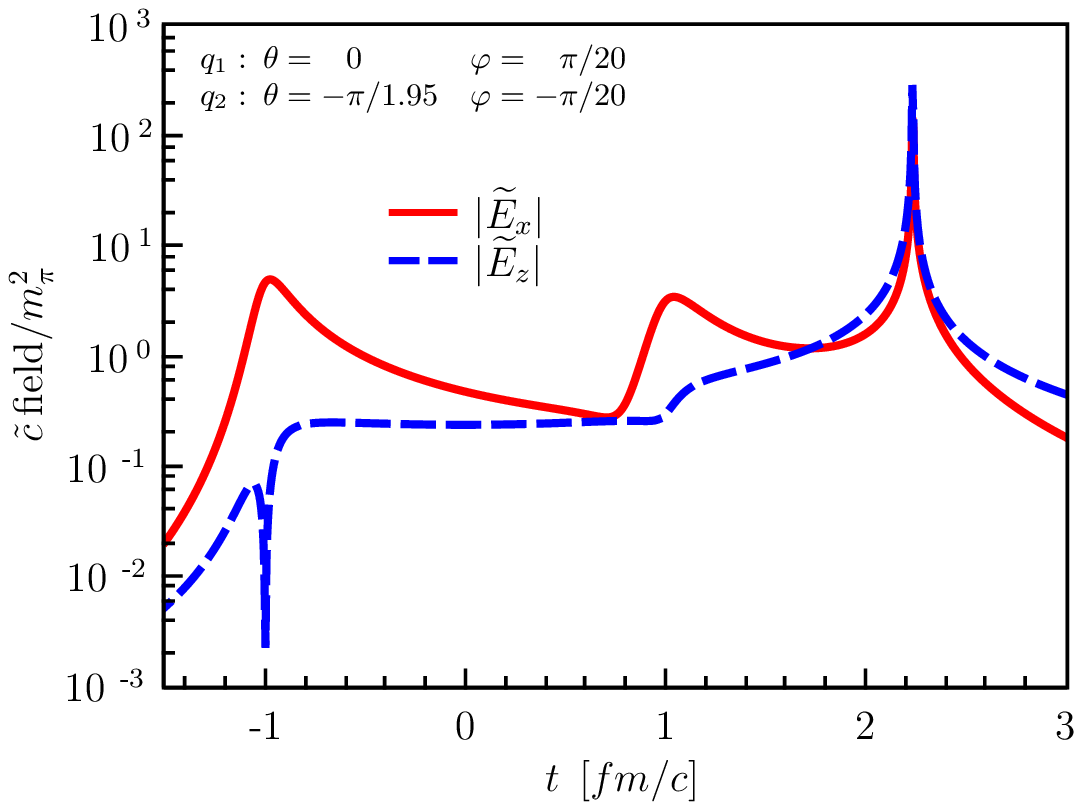}
\vspace{-3mm}
\begin{picture}(0,0)(0,0)
\put(-270,145){\small({\bf a})}
\put(-35,145){\small({\bf b})}
\end{picture}
 \caption{(Color online) Modulus of the two components  of the
chromoelectric field strength $| \widetilde E_x |$ (solid line)
and $|\widetilde E_z|$  (dashed line) for two moving charges with
parallel (a) and orthogonal (b) color charges
in the initial time. }
 \label{f5}
\end{center}
\end{figure*}
\begin{figure*}
\begin{center}
\includegraphics[width =0.45\textwidth, clip = true] {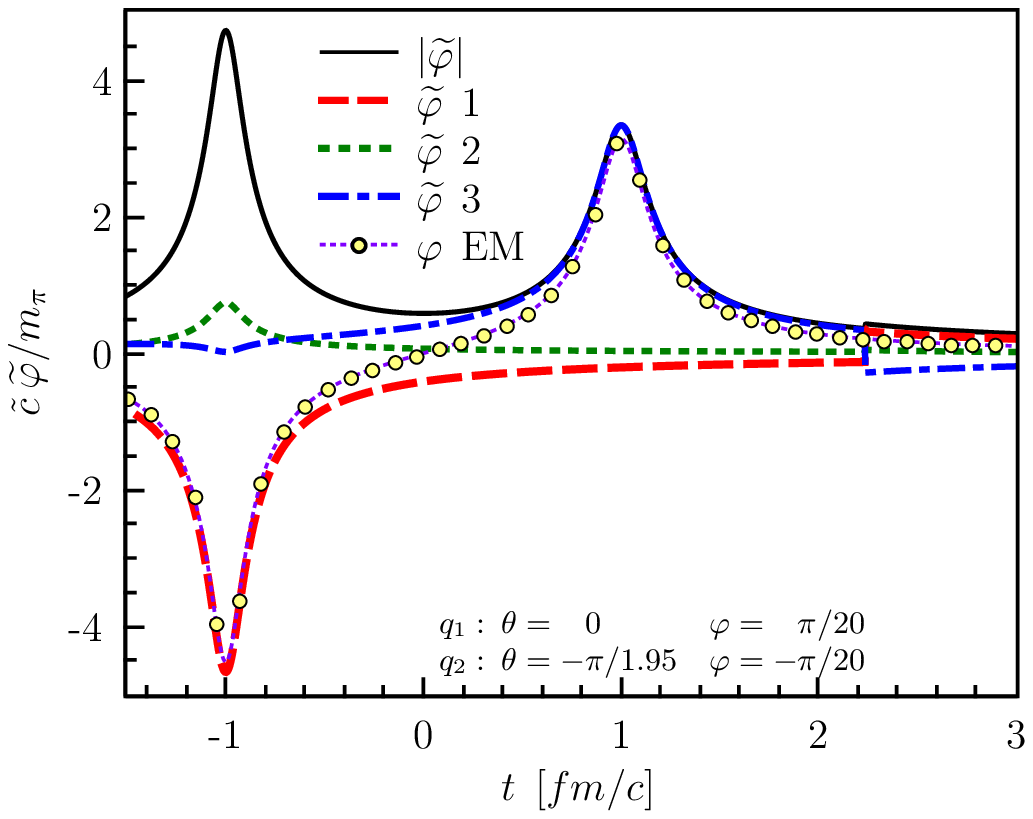}
\hspace{0.5cm}
\includegraphics[width =0.45\textwidth, clip = true] {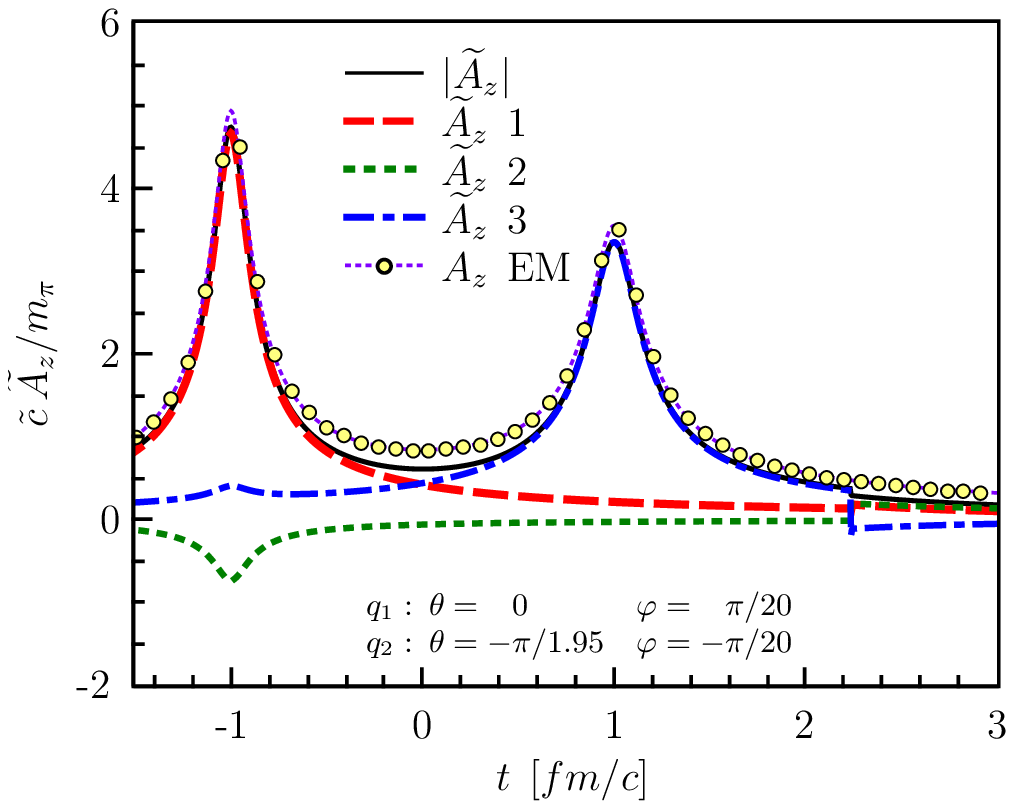}
\vspace{-5mm}
\begin{picture}(0,0)(0,0)
\put(-285,155){\small({\bf a})}
\put(-35,155){\small({\bf b})}
\end{picture}
\caption{(Color online) Non-Abelian configuration of
color charges (see text). The three components of $\widetilde \phi
$  (a) and $\widetilde A_z $  (b) are shown for
the initially orthogonal charge vectors. The notation is the same
as in Fig.~\ref{f3}. }
 \label{f6}
\end{center}
\end{figure*}
\begin{figure*}
\begin{center}
\includegraphics[width=0.45\textwidth,clip=true] {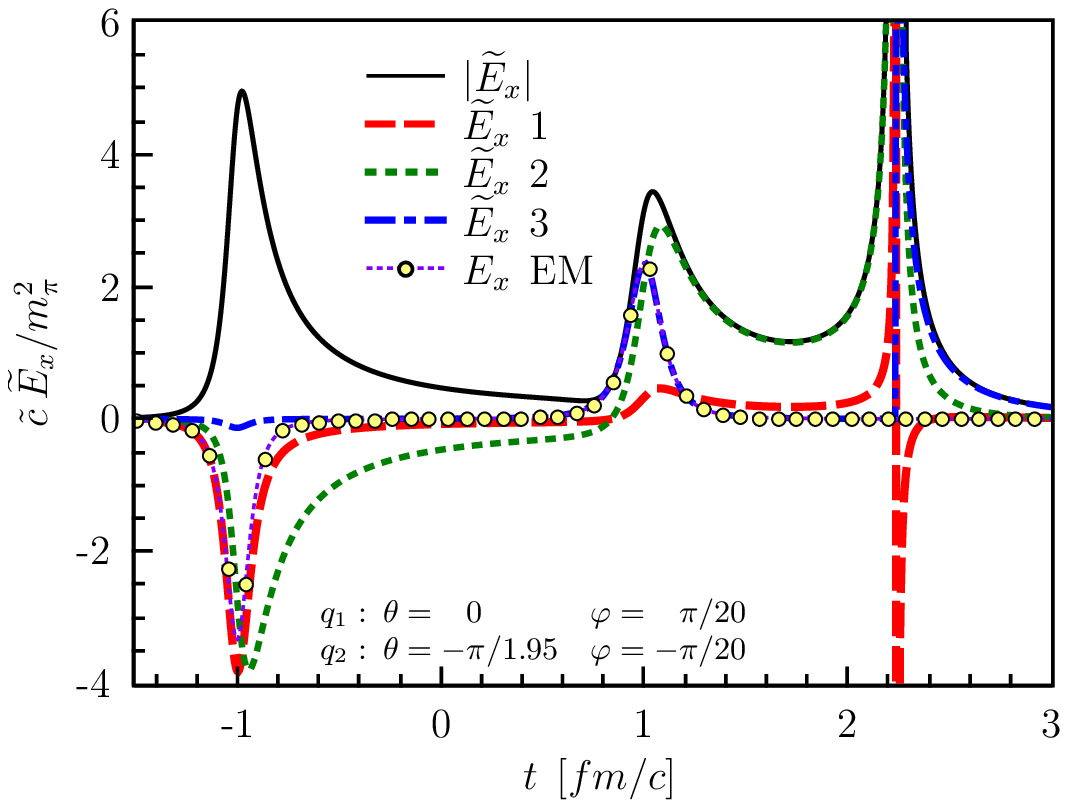}
\hspace{0.5cm}
\includegraphics[width =0.45\textwidth, clip = true] {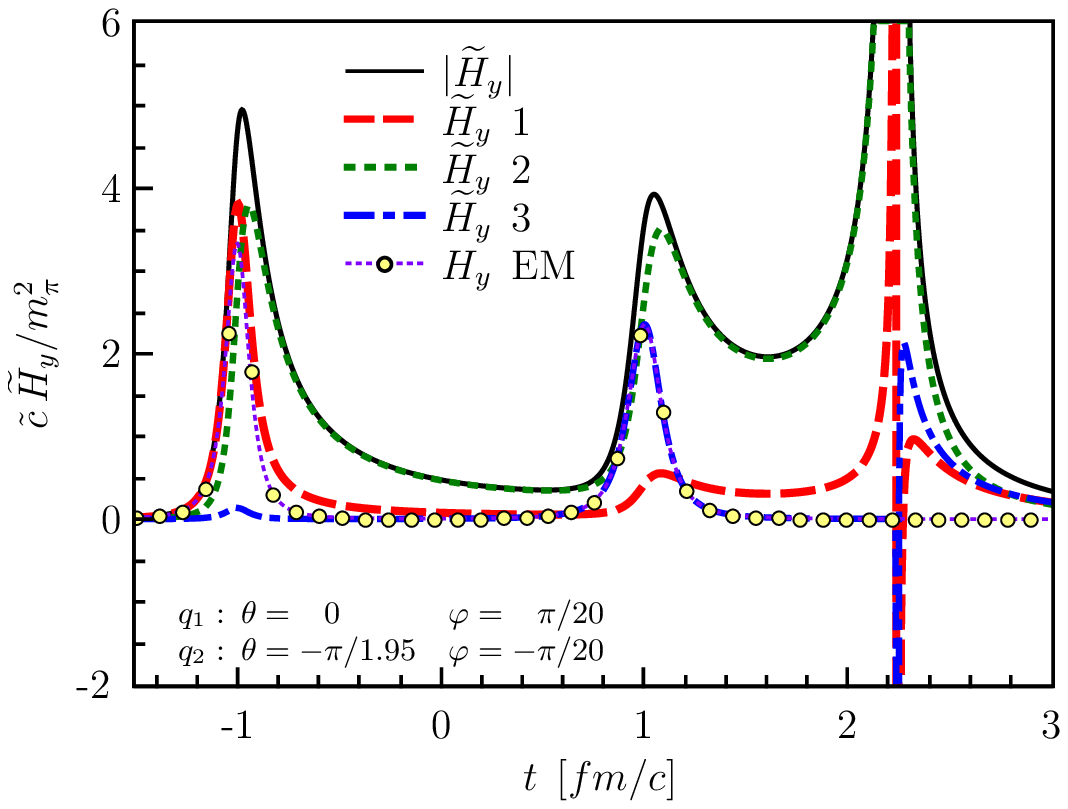}
\vspace{-3mm}
\begin{picture}(0,0)(0,0)
\put(-285,150){\small({\bf a})}
\put(-30,150){\small({\bf b})}
\end{picture}
\caption{(Color online) Non-Abelian configuration of
color charges (see text).  The three components of chromoelectric
$ \widetilde E $ (a) and chromomagnetic $ \widetilde H $
(b) fields are shown for the initially orthogonal charge
vectors. The notation is the same as in Fig.~\ref{f4}. }
 \label{f7}
\end{center}
\end{figure*}

The procedure for obtaining an approximate solution is given in
the Appendix. To find the strength of the chromoelectric and
chromomagnetic fields, it is also required to know the derivatives
of the color charge vectors with respect to the retarded time $d\widetilde
P(t')/dt'$, $d\widetilde Q(t')/dt'$. They are computed by the
explicit formulas (\ref{32}).

As an example, let us consider the field created by  two
relativistic particles moving with velocities: $v=1-\dmn{2}{-3}$
and $|w|=1-\dmn{1}{-3}$. The energy-mass ratio for the first
particle is ${\cal E}/m\simeq$ 16, and for the second particle is
${\cal E}/m\simeq$ 22. Let us take the coordinates of the
observation point as $\vf{r}_0=(2,0,1)$, i.e., $x=2$ fm, $z=1$ fm.
For comparison, we consider also the field created by particles
with electric charge $\pm e$ of the same interaction strength as
the color charges ($e^2/(4\pi)=g^2/(4 \pi)=0.3$) moving with the
same velocities. The potentials for the scalar $ \widetilde \varphi $
and $z$ component of the vector $ \widetilde A_z $ fields of color
particles are presented in Fig.~\ref{f3} where the initial angles
in the color space are determined as $\theta=\pi/1.95 $,
$\phi=\pi/20$ for the first particle and $\theta=-\pi/1.95$,
$\phi=-\pi/20$ for the second one. In this case the color charge
of the first particle, for example, will be $\widetilde
P=(P_1,P_2,P_3)$, with $P_1=\sin\theta\cos\phi$,
$P_2=\sin\theta\sin\phi$, $P_3=\cos\theta$. This configuration of
color charges  $\widetilde P$, $\widetilde Q$  corresponds to
almost oppositely directed charges at the initial stage. The
dashed lines in this figure show the three color components of the
scalar potential [Fig.~\ref{f3}(a)] and the third component of the
vector potential [Fig.~\ref{f3}(b)]. The open circles are plotted for the
potentials corresponding to the enhanced electrodynamic coupling.
The modulus of the scalar $|\widetilde\varphi|$ and vector
$|\widetilde A_z|$ potential are displayed by the solid lines in
Fig.~\ref{f3}. We recall again that these variables are not
directly observable. The selected observation point
$\vf{r}_0=(2,0,1)$ introduces some asymmetry reminding us of the
asymmetry in the transverse plane for peripheral nucleus-nucleus
collisions which lead to some dominant component~\cite{Ton}.

According to the choice of the geometry, we see  that there
are three maxima in the evolution of the chromo field,
Fig.~\ref{f4}. The first
maximum corresponds to the passage of the first particle at the
closest distance to the observation point, then the second
maximum corresponds to the passage of the second charge and
a late third one for which there is no analogy in the
electromagnetic field (the dotted line is flat in this time
interval). The first two bumps are located symmetrically with
respect to the meeting point $t=0$.
For the selected configuration of color charges, one of the
components of the potentials, shown by the dashed lines,
dominates and almost coincides with the appropriate modulus of
vectors in the color space (solid line).  The meeting point
 seen by an observer is
located at a distance of $R=(x^2+z^2)^{1/2}=\sqrt{5}\sim 2.24$
from $t=0$.  It is seen that there is a noticeable
difference between the scalar and vector potentials as compared to
the electrodynamic case (open circles) at the appropriate time.
As discussed in the Appendix, the rotation of color charges is
described by Eqs.~(\ref{35})--(\ref{39}). The effective rotation
frequency in the neighborhood of the meeting point $t\to 0$ is
$\omega''=\alpha_q / ln |t|$, i.e., the color charges are rotating
infinitely fast near the meeting point.

The time dependence  of the strength components of the
chromoelectric $\widetilde E_x$ [Fig.~\ref{f4}(a)] and chromomagnetic
$\widetilde H_y$ [Fig.~\ref{f4}(b)] fields is shown in Fig.~\ref{f4} by
the three dashed lines; the solid lines correspond to the modulus
of the chromofields $|\widetilde E_x|$ and $|\widetilde H_y|$. We
have cut the singular peaks at some threshold and thus the lines
look somewhat irregular. In full agreement with Eqs.~(\ref{25}),
the $E_x$ and $H_y$ components are dominating. In both cases
 two maxima (minima) - caused by passing the color charges in the
vicinity of the observation point - are clearly visible.  Note that
here (and in what follows) the color field strength is plotted in
dimensionless units where $\widetilde c$ is the color charge of the
appropriate field component and $m_\pi$ is the pion mass. The charge
velocities considered roughly correspond to the RHIC energy where
the maximal electromagnetic field $eH_y/m_\pi^2$ reaches a few
units~\cite{Ton}. This value is essentially smaller than those in
the color charge case (see Fig.~\ref{f4}). It is seen that for the
color charge configuration considered the chromoelectric and
chromomagnetic field are  quite similar to the field in the case of
enhanced ($\alpha_e=$0.3) electrodynamics. Some difference in the height of the first
two maxima are caused  by different velocities of color charges.  A
significant difference at the third maximum is due to the arrival of
a signal from the meeting point of particles to the observation
point, where there is a noticeable additional contribution of
chromoelectric and chromomagnetic fields associated with the
temporal change of the particle color charges.  This extra
enhancement may be considered as a manifestation of the color
vector rotation in the evolution of the color field strength named
as the effect of the "color charge glow" \ . The color glow effect  is
not  an artifact of the approximation  but results from the pure
non-Abelian term proportional to $\widetilde D$ in Eq.~(\ref{25})
 as a distinct
color wave disturbance arising due to the finite retardation time.

The modulus of the transverse $|\widetilde E_x|=\sqrt{\widetilde
E_x^2}$ and longitudinal $|\widetilde E_z|=\sqrt{\widetilde E_z^2
}$ component of the chromoelectric field is shown in Fig.~\ref{f5}
for two configurations of the initial color charges. The
longitudinal component is strongly suppressed, as it should be due
to relativistic effects, but the signal from the meeting point of
the particles leads to almost equal contributions. It is
noteworthy that the change of the initial configuration of color
charges from parallel to orthogonal  (cf. (a) and (b) panels in
Fig.~\ref{f5}) does not change the evolution in the absolute
values of the chromoelectric field though the signs of the color
$E_x, E_z$ components are different and change with time. It is
also important to note that every collision noticeably changes the
position of color charges in the color space  but we do not present
these scattering data here.

Characteristics for the initially almost orthogonal color
charge vectors  are presented in detail in Figs.~\ref{f6} and
\ref{f7}. These results are obtained for particles with color
charges  defined by the following angles in  color space:
$\theta=0$, $\phi=\pi/20$ for the first particle
$\theta=-\pi/1.95$, $\phi=-\pi/20$ for the second one. We shall
henceforth refer to this case as non-Abelian. The notation in
these figures is identical to that in Figs.~\ref{f3} and \ref{f4}.
The main result here is that in this case there is no dominant
component, as in the case of mutually opposite charges considered
above, but two preferred directions in color space are
significant, as evidenced by the corresponding maxima shown by the
dashed lines in Figs.~\ref{f6} and \ref{f7}.  The third bump
at $t=2.24$ is again the manifestation of the color charge glow
effect.

Some comments with respect to previous studies - addressing the
color rotation in the encounters of color charges - are in order:
In Ref. ~\cite{KR97} the authors have solved the classical QCD
equations of motion in perturbation theory up to order $g^3$
within light-cone variables (using the Mueller gauge
transformation~\cite{Mue94}) and assuming  that the color charges
move with the speed of light ($v=1$) and the color interaction is
switched on at $T=-\infty$. Their focus has been on the
computation of 'soft' gluons in the initial phase of
ultrarelativistic nucleus-nucleus collisions in the forward
light-cone, in particular the gluon number, the gluon energy and
their multiplicity distributions. Though the basic equations
(\ref{12}) and (\ref{13}) are the same, our model considers charges
moving with velocities $v<1$, subdivides the
scattering process at three different stages (see Appendix) and
takes into account the finite retardation time at each stage.
As demonstrated above in Figs.~\ref{f4} and~\ref{f6}, we are
interested explicitly in the evolution of chromoelectric and
chromomagnetic fields where the new color glow effect is observed most
clearly. Since the authors of Ref.~\cite{KR97} compute the gluon
field obtained by the Weizs\"acker-Williams transformation of the
potential and consider only global properties such as energy,
number, and multiplicity distributions of gluons, a possible
contribution of the color charge glow is hard to discriminate.

Furthermore, the space-time evolution of the classical gluon fields
was investigated before in Ref.~\cite{MMR98} within a rather similar
non-Abelian model. This model, applied to the collision of two
nuclei, exhibits a very complicated field structure associated with
instabilities. In this picture, it is hardly possible to disentangle
such a particular mode as the late effect of the color glow, which
in principle could have left its traces also in their calculations.
However, an additional investigation for simple colliding systems,
as performed here, is mandatory to clearly pin down this
phenomenon.

\section{The field of a color charge and a color dipole}
\begin{figure}
\includegraphics[width = 7.cm, clip = true] {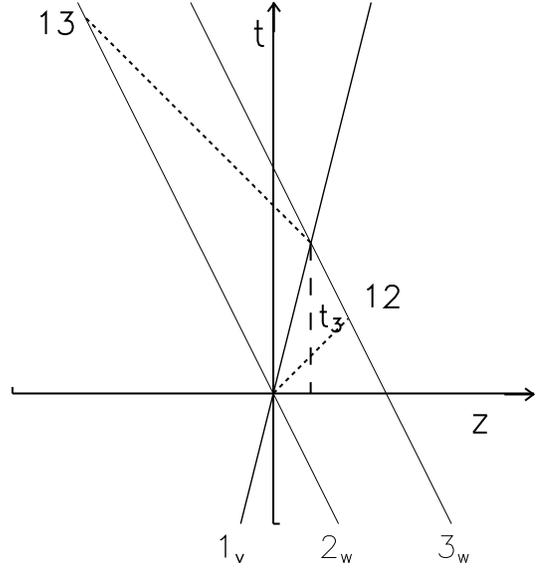}
\caption{Scheme of the meeting of a color particle and a dipole. The
particle trajectories are plotted by the solid lines, the dotted
lines are light-cone lines corresponding to the arrival of the
signal from the particle meeting point.  The arrival of the signal
is marked by the points $12$ and $13$ (see text). The time $ t_3 $
is the moment when the first and the third particle meet each
other. } \label{f8}
\end{figure}

As in the previous section, let us assume that there is a particle
with color charge $\widetilde P$ that moves along the $z$ axis with
velocity $v$. We denote it as the first particle. Suppose that the
dipole made up by the second and third particle with charges
$\widetilde Q_2$ and $\widetilde Q_3$ is aligned along the $z$ axis
and moves in the opposite direction with velocity $w$. We denote the
distance between the charges in the dipole as $\delta_w$ in its rest
frame which is taken as $1$ fm. In the laboratory frame the charges
are located closer to each other $\Delta_w'=(1-w ^2)^{1/2}\delta_w$
due to the Lorentz contraction. Let the first and second particle
encounter at zero time before the time $t_3=\frat{\delta_w'}{v+w}$
of the second meeting between the first and third particle (see
Fig.~\ref{f8}). Then the trajectories of the first, second and third
charges  are defined as $z_1=vt$, $z_2=-wt$, $z_3=-wt+\delta_w '$.
Similarly to Eq.~(\ref{30}), let us adopt an approximate solution of
the Yang-Mills equations in the form
\begin{eqnarray}
\label{42} \widetilde\varphi&=&[\varphi_1 \widetilde P]_{t'}+
[\varphi_2\widetilde Q_2]_{t'}+[\varphi_3 \widetilde
Q_3]_{t'}~,\nonumber \\ [-0.1cm] \\[-0.25cm] \widetilde
A_z&=&v~[\varphi_1
\widetilde P]_{t'}-w~[\varphi_2 \widetilde Q_2]_{t'} -w~[\varphi_3
\widetilde Q_3]_{t'}~. \nonumber
\end{eqnarray}
The potentials at the location point of charges can be found  in the
same way as in the case of two color charges; this leads to the
compatibility conditions for each of the charges, see
Eq.~(\ref{23})
\begin{eqnarray}
\label{43} \dot{\widetilde P}&=&\alpha_g~\frat{1+v~w}{|z_1-z_2 |}~
\widetilde Q_2 (t-t^{**}_{12}) \times \widetilde P +
\alpha_g~\frat{1+vw}{|z_1-z_3|}~\nonumber \\  &\times&\widetilde
Q_3(t-t_{13}^{***}) \times \widetilde P~, \nonumber \\
\dot{\widetilde Q}_2&=&\alpha_g~\frat{1+v~w}{|z_1-z_2|}~
\widetilde P(t-t^{*}_ {21}) \times \widetilde Q_2 +
\alpha_g~\frat{1-w^2}{|z_2-z_3|}~\nonumber \\ &\times&\widetilde
Q_3(t-t^{***}_{23})\times \widetilde
Q_2~,\\
\dot{\widetilde Q}_3&=&\alpha_g~\frat{1+v~w}{|z_1-z_3|}~
\widetilde P(t-t^{*}_{31})\times \widetilde Q_3 + \alpha_g
~\frat{1-w^2}{|z_2-z_3|}~\nonumber \\ &\times&\widetilde
Q_2(t-t^{**}_{32}) \times \widetilde Q_3~, \nonumber
\end{eqnarray}
with $t^{*}_{31}=-\frat{v+w}{1+v}(t-t_3)$,
$t^{***}_{13}=-\frat{v+w}{1+w}(t-t_3)$ for $t_3<t$  and
$t^{**}_{32}=\frat{v+w}{1+w}~t_3$,
$t^{***}_{23}=\frat{v+w}{1-w}t_3$ for $t<t_3$ (with the
substitution $v,w \to-v,-w$). The retardation times are determined
similarly to the case of two color charges, where, in particular,
the retarded times $t^*_{21}$ and $t^{**}_{12}$ are given (see the
relevant scheme in Fig.~\ref{f2}). An interesting peculiarity of
the resulting system of Eqs.~(\ref{43}) is  strong suppression of
the contributions of charges 2 and 3  flying in the same
direction, which enter into the equation with the Lorentz factor
$1-w^2$. Thus, in the ultrarelativistic case the mutual influence
of color charges flying in the same direction may be ignored and
the system can be considered as "frozen"{\phantom {.}}. This is
true even in the cases where the system has the size of a nucleus.

We are interested now in the particular case when the color dipole
charges in
the initial state  are opposite, i.e. $\widetilde Q_3=-\widetilde Q_2$.
The prescription for obtaining an approximate solution in this case is
given in
the Appendix.
\begin{figure*}
\begin{center}
\includegraphics[width =0.45\textwidth, clip = true] {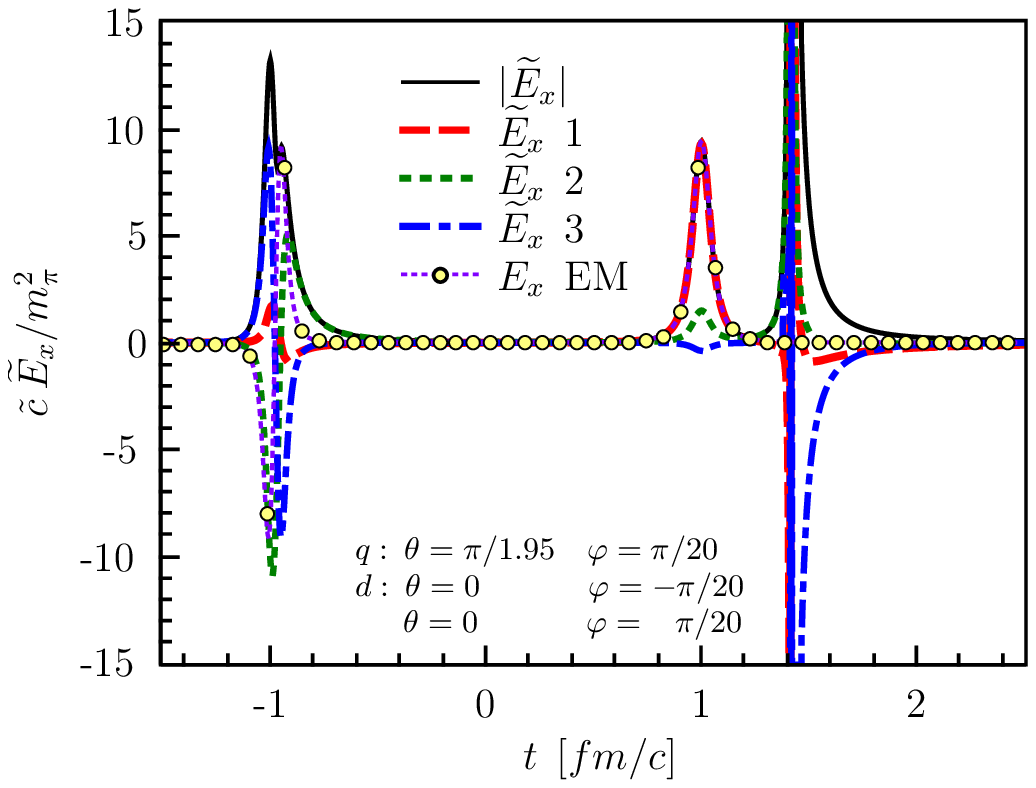}
\hspace{0.5cm}
\includegraphics[width =0.45\textwidth, clip = true] {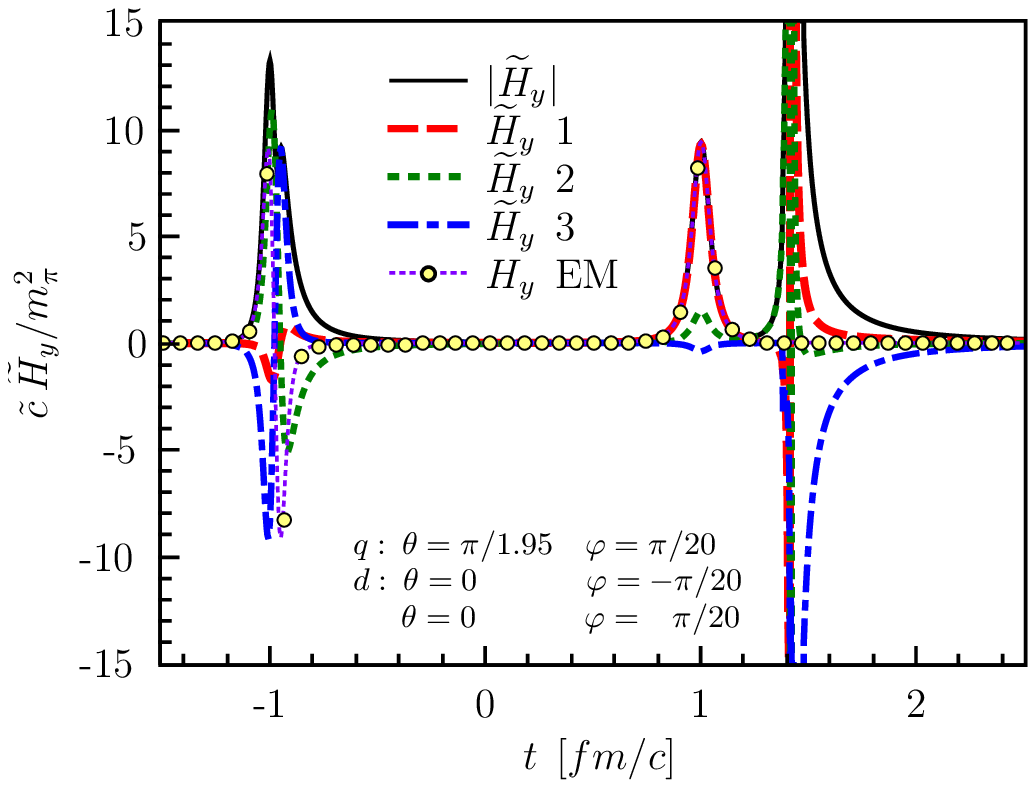}
\vspace{-2.5mm}
\begin{picture}(0,0)(0,0)
\put(-285,150){\small({\bf a})}
\put(-30,150){\small({\bf b})}
\end{picture}
\caption{(Color online) Chromoelectric  $
\widetilde E $ (a) and chromomagnetic $ \widetilde H_y $
(b) fields for the charge-dipole case. The notation is
the same as in Fig.~\ref{f4}. } \label{f9}
\end{center}
\end{figure*}


The strength of the chromoelectric  and chromomagnetic fields
generated by particles moving with the velocity $v=1-2 \ 10^{-2}$
and by the dipole with the velocity $|w|=1-1 \ 10^{-2}$  is
demonstrated in Fig. \ref{f9}. The color charges of the particles
at the initial time are determined by the following angles in the
color space: $\theta=\pi/1.95$, $\phi=\pi/20$ for the first
particle, $\theta=0$, $\phi=-\pi/20$ for the second one and with
the opposite angles for the third particle. This configuration is
denoted as "non-Abelian"\ . We do not show explicit data for the
potentials, as they provide little information. It is seen that
generally the two first maxima of the field strength generated by
color charges in the vicinity of the observation point are
reasonably reproduced by the Coulomb-like solution (dotted lines
in Fig.~\ref{f9}). However, the first maximum formed in passing
the dipole is not smooth but has an "up-down jump" (or zig zag)
shape. This structure is caused by the opposite color charges
forming the dipole.

The observation point is located now in another place $x=1$, $z=1$
fm  as compared to the case of the meeting of two particles at
$x=2$, so the signal of the meeting of the color charge and the
dipole arrives  at a time about
$\tau=(x^2+z^2)^{1/2}/c=\sqrt{2}\sim 1.41$. Here, at the meeting
point the Coulomb-like solution predicts noticeably narrower
distributions over the color field strength due to the so-called
color charge glow, which, as noted above, results from the color
charge interaction through the time dependence of the color
vectors $\widetilde D$ (the last term in Eqs.~(\ref{25})).

\section{The field of two color dipoles}
\begin{figure}
\includegraphics[width = 7.cm, clip = true] {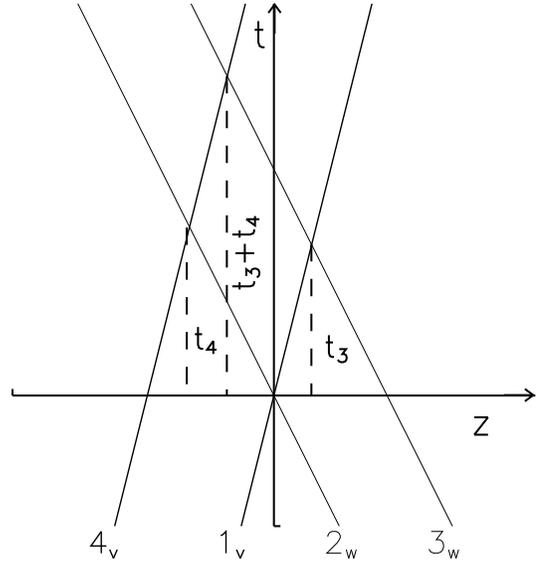}
\caption{Scheme of the meetings of two color dipoles. The solid lines
are the particle trajectories.  The meeting points of the
particles are marked by $t_3$ for the  particles 1-3, $t_4$ for
the  particles 4-2 and $t_3+t_4$ for the particles 4-3.}
\label{f10}
\end{figure}

Let us finally consider the field created by  two color dipoles.
The first dipole is formed by the first and fourth particle with
charges $\widetilde P_1$ and $\widetilde P_4$, respectively. The
second dipole is made up of the second and third particle with
charges $\widetilde Q_2$ and $\widetilde Q_3$. At the initial time
the color charge  of each dipole is neutral: $\widetilde
P_1=-\widetilde P_4$, $\widetilde Q_2=-\widetilde Q_3$. Particles
of the first aligned dipole move along the $z$ axis with
velocity $v$ and the second aligned dipole moves towards them with
 velocity $w$. Let us denote the distance between the charges
in the first and second dipole in their rest system  as $\delta_v$
and $\delta_w$, respectively. In the laboratory frame, these
distances are contracted $\delta_v'=(1-v^2)^{1/2}\delta_v$ and
$\delta_w'=(1-w^2)^{1/2}\delta_w$. Again, it is convenient to
introduce the time scales
$$t_3=\frat{\delta_w'}{v+w}~,~~t_4=\frat{\delta_v'}{v+w}~,$$
when the first particle  meets the third one and the fourth meets the
second one, respectively (see Fig.~\ref{f10}). At the time
$t_3 + t_4 $  the fourth particle meets the third  one.
\begin{figure*}
\begin{center}
\includegraphics[width =0.45\textwidth, clip = true] {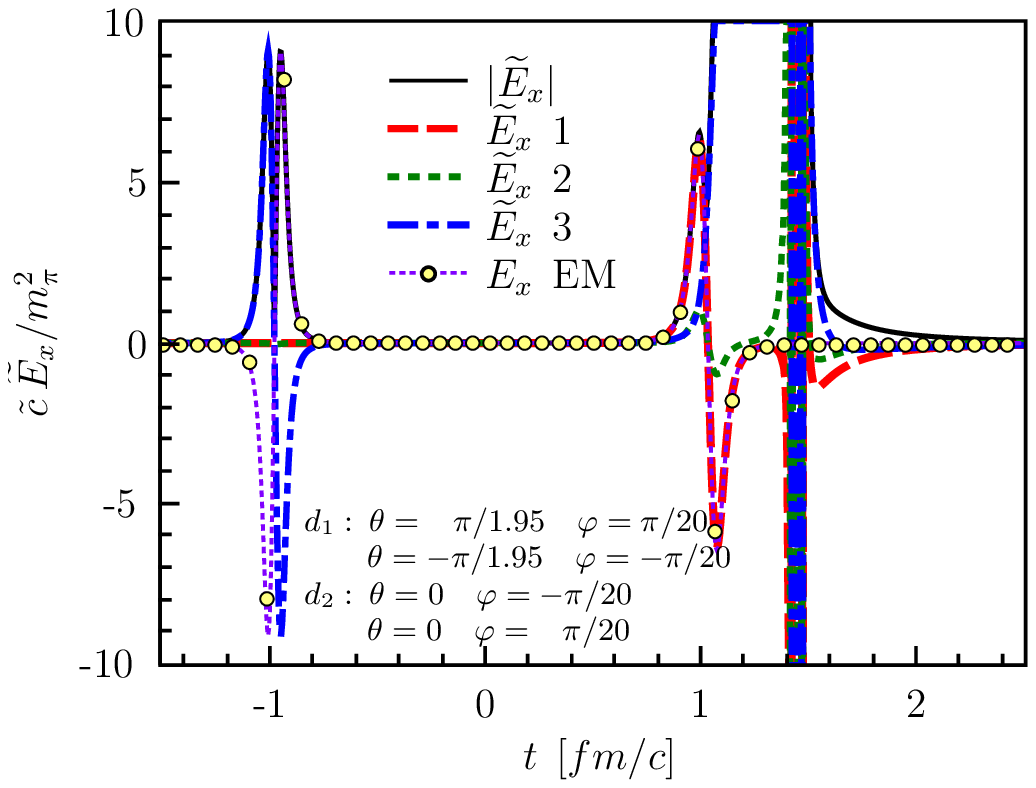}
\hspace{0.5cm}
\includegraphics[width =0.45\textwidth, clip = true] {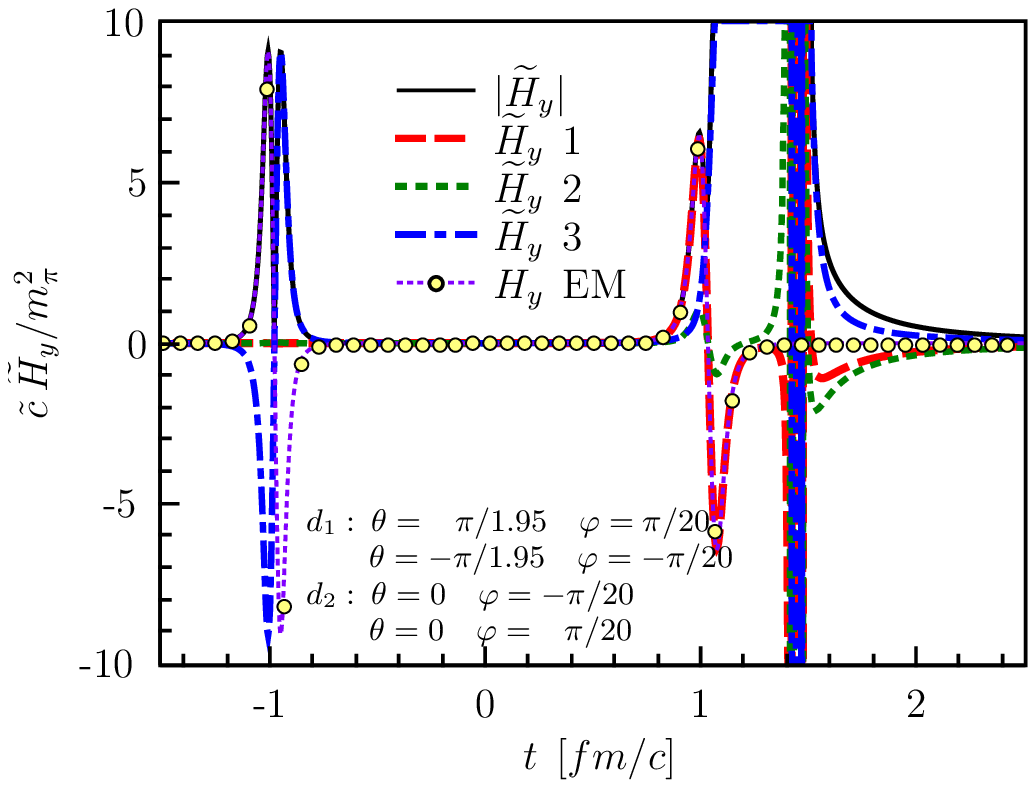}
\vspace{-2.5mm}
\begin{picture}(0,0)(0,0)
\put(-285,150){\small({\bf a})}
\put(-30,150){\small({\bf b})}
\end{picture}
\caption{(Color online) The three components of $
\widetilde E_x $ (a) and $ \widetilde H_y $  (b)
for the dipole-dipole case. The notation is similar to
Fig.~\ref{f4}. } \label{f11}
\end{center}
\end{figure*}
\begin{figure*}
\begin{center}
\includegraphics[width =0.45\textwidth, clip = true] {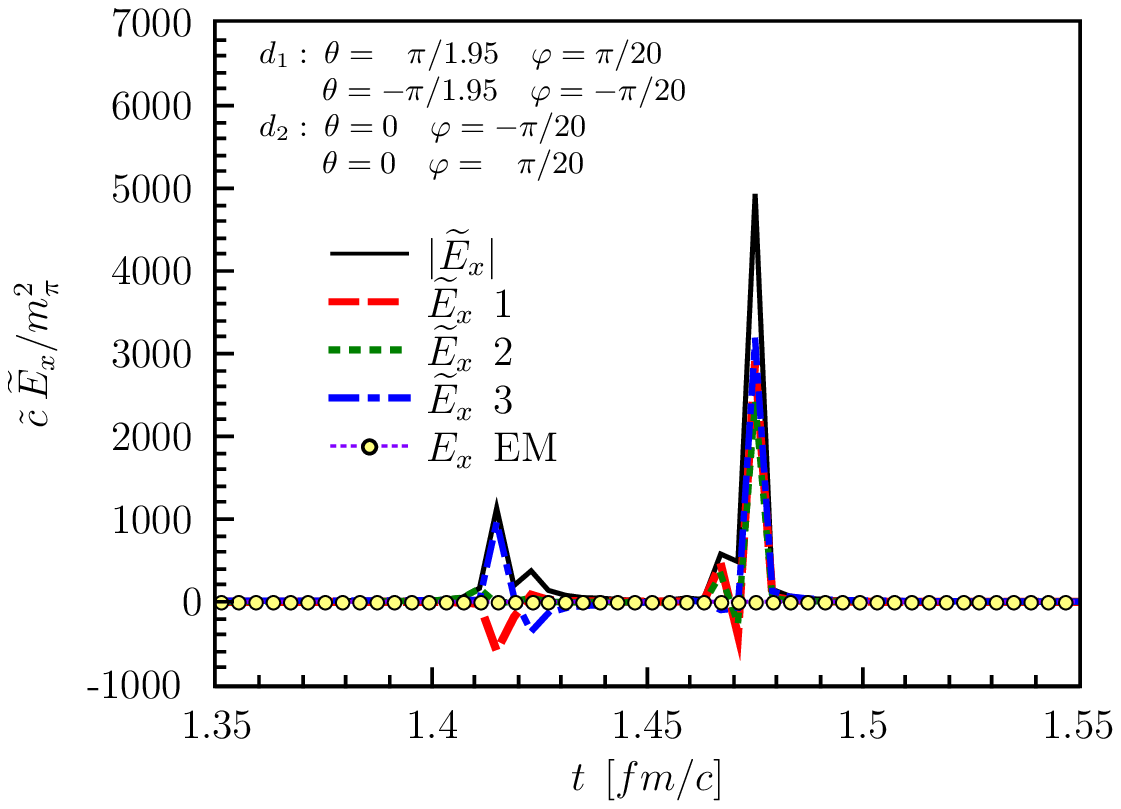}
\hspace{0.5cm}
\includegraphics[width =0.45\textwidth, clip = true] {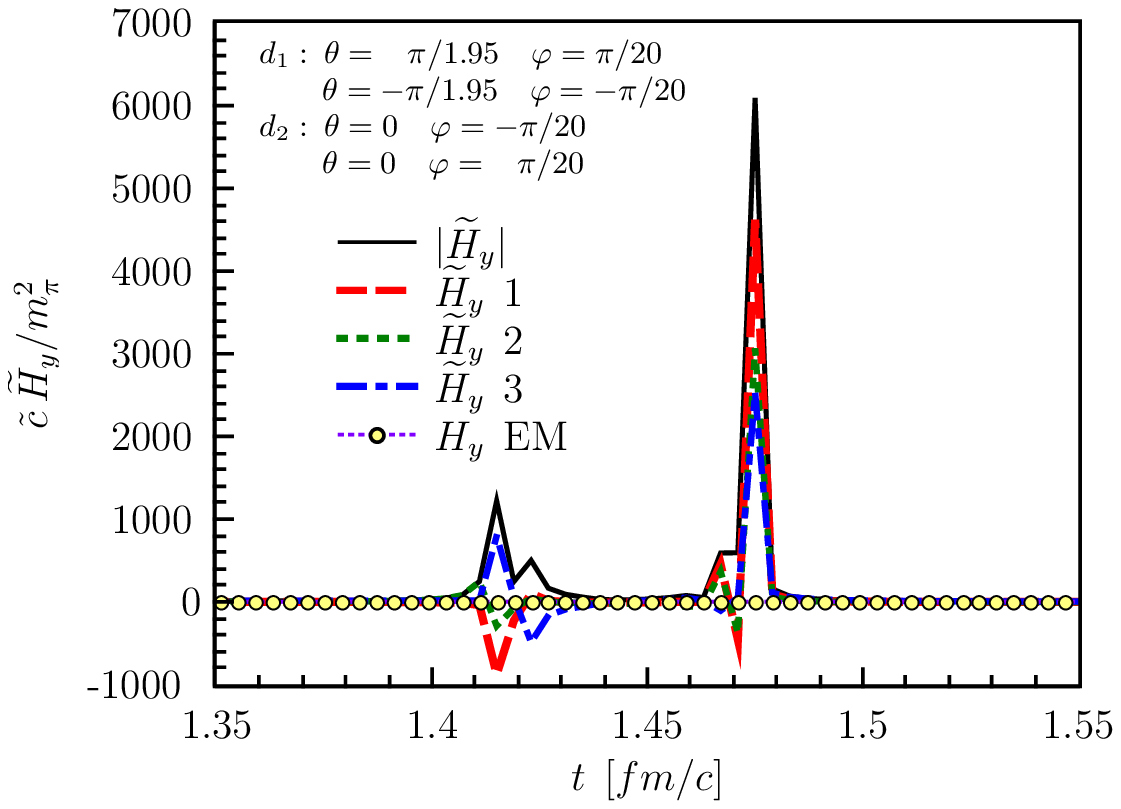}
\vspace{-2.5mm}
\begin{picture}(0,0)(0,0)
\put(-295,140){\small({\bf a})}
\put(-40,140){\small({\bf b})}
\end{picture}
\caption{(Color online) The same as in
Fig.~\ref{f11} but for the zoomed region near the meeting point. }
\label{f12}
\end{center}
\end{figure*}
Suppose for convenience that the meeting of the first and second
particle occurs at time zero. Then the trajectories of charges
with numbers one, two, three and four are defined as $ z_1 = vt $,
$z_2=-wt$, $z_3=-wt+\delta_w'$, $z_4=vt-\delta_v'$.
 In this case  the ansatz for the superposition of approximate solutions
has the form
\begin{eqnarray}
\label{49} \widetilde\varphi&=&[\varphi_1 \widetilde
P_1]_{t'}+[\varphi_2 \widetilde Q_2]_{t'} +[\varphi_3 \widetilde
Q_3]_{t'}+[\varphi_4 \widetilde P_4]_{t'}~,
\nonumber \\[-.1 cm]
\\[-.2cm]
\widetilde A_z&=&v~[\varphi_1 \widetilde P_1]_{t'}- w~[\varphi_2
\widetilde Q_2]_{t'}-w~[\varphi_3 \widetilde Q_3]_{t '}
+v~[\varphi_4 \widetilde P_4]_{t'}~. \nonumber
\end{eqnarray}
Similarly to the previous sections  the compatibility condition
(see Eq.~(\ref{23})) for each  charge can be written as follows~:
\begin{eqnarray}
\label{50} &&\dot{\widetilde
P}_1=\alpha_g~\frat{1+v~w}{|z_1-z_2|}~ \widetilde
Q_2(t-t^{**}_{12}) \times \widetilde P_1
+\alpha_g~\frat{1+vw}{|z_1-z_3|}~ \nonumber \\ &\times& \widetilde
Q_3(t-t_{13}^{***})\times \widetilde P_1
+\alpha_g~\frat{1-v^2}{|z_1-z_4|}~
\widetilde P_4(t-t^{IV}_{14})\times \widetilde P_1, \nonumber\\
&&\dot{\widetilde Q}_2=\alpha_g~\frat{1+v~w}{|z_1-z_2|}
~\widetilde P_1(t-t^{*}_{21}) \times \widetilde Q_2
+\alpha_g~\frat{1-w^2}{|z_2-z_3|}~ \nonumber
\\ &\times& \widetilde Q_3(t-t^{***}_{23})\times \widetilde Q_2
+\alpha_g~\frat{1+v~w}{|z_2-z_4|}~
\widetilde P_4(t-t^{IV}_ {24})\times \widetilde Q_2, \nonumber\\[-0.1cm]
\\[-.2cm]
&&\dot{\widetilde Q}_3=\alpha_g~\frat{1+v~w}{|z_1-z_3|}~
\widetilde P_1(t-t^{*}_{31}) \times \widetilde Q_3 +
\alpha_g~\frat{1-w^2}{|z_2-z_3|}~ \nonumber \\ &\times& \widetilde
Q_2(t-t^{**}_{32})\times \widetilde Q_3
+\alpha_g~\frat{1+v~w}{|z_3-z_4|}~
\widetilde P_4(t-t^{IV}_{34}) \times \widetilde Q_3, \nonumber\\
&&\dot{\widetilde P}_4=\alpha_g~\frat{1-v^2}{|z_1-z_4|}
~\widetilde P_1(t-t^{*}_{41}) \times \widetilde P_4
+\alpha_g~\frat{1+v~w}{|z_4-z_2|}~  \nonumber \\ &\times&
\widetilde Q_2(t-t^{**}_{42})\times \widetilde P_4
+\alpha_g~\frat{1+v~w}{|z_3-z_4|}~ \widetilde Q_3(t-t^{***}_{43})
\times \widetilde P_4, \nonumber
\end{eqnarray}
where the different times are defined as
$t^{**}_{42}=-\frat{v+w}{1-w}(t-t_4)$,
$t^{IV}_{24}=-\frat{v+w}{1-v}(t-t_4)$ for $t<t_4$ ( for $t_4 <t$,
the substitution should be made $v,w\to$ $-v$, $-w$); \
$t^{***}_{43}=-\frat{v+w}{1-v}(t-t_3-t_4)$,
$t^{IV}_{34}=-\frat{v+w}{1-w}~(t-t_3-t_4)$ for $t<t_3+t_4$
 (for $t_3+t_4<t$, the substitution should be made $v,w\to$ $-v$, $-w$);
$t^{*}_{41}=-\frat{v+w}{1-v}t_4$ and $t^{IV}_{14}=-\frat{v+w}{1+v}t_4$.
In  the Appendix we present  the procedure for obtaining an
approximate solution for two color dipoles.

In Fig.~\ref{f11}, the strength of chromoelectric  and
chromomagnetic fields  created by two dipoles with the velocities
$v=1-2 \ 10^{-2}$ and $|w|=1-1 \ 10^{-2}$ is demonstrated. The
color charges of the particles at the initial time are determined
by the following angles in the color space: $\theta=\pi/1.95$,
$\phi=\pi/20$ for the first particle and $\theta=0$,
$\phi=-\pi/20$ for the second one and with the opposite angles for
the third and fourth particle.  As is seen (by the dotted lines in
Fig.~\ref{f11}), the two first maxima are reproduced  by the
Coulomb-like solution including the up-down jump effect for both
maxima, noted above in Sec. V. The observation point has the
coordinates $x=1$ fm and $z=1$ fm, so the signal on the meeting of
the color dipoles arrives at the time about
$\tau=(x^2+z^2)^{1/2}/c=\sqrt{2}\sim 1.41$. In fact, the whole
area from the second maximum until the meeting point is filled by
the chromoelectric and chromomagnetic fields of noticeable
strength and near $\tau=1.41$ this looks like a band. To clarify
the band structure, we zoom into the region near the meeting point in
Fig.~\ref{f12}. It is seen that  this area consists of four
maxima of huge intensity, which in dynamical systems can result in
large local color fluctuations. This shining of the color charge
glow is mainly due to color charge interactions through the time
dependence of the color vectors $\widetilde D$ [the last term in
Eqs.~(\ref{25})], which is certainly not reproduced by the Coulomb
solution with the enhanced coupling. It is noteworthy that the
color charge glow effect is getting even larger for more
complicated systems.

\section{Conclusions}
In this study, we have considered several elementary configurations
of relativistic partons with  non-Abelian charges for the
SU(2) group in the classical limit. It is shown that as in the
case of non-relativistic particles, the system generally shows
Coulomb-like features, and the analogy is convincingly supported
by a comparison with electrodynamics for the same coupling
strength. In distinction, we find an additional strength of
chromoelectric and chromomagnetic fields close to the meeting of
particles, which is caused by the explicit time dependence of the
color charge vectors. In the chosen gauge the interaction of the
color charges results in the rotation of the color vectors, which
becomes very fast close to the meeting point of two particles. The
chromo fields are stronger by about an order of magnitude than the
corresponding electromagnetic fields created by a moving source at
similar conditions. The longitudinal and transverse field
components of this signal are of the same order of magnitude in
contrast to the longitudinally compressed field (due to the
Lorentz contraction). Changing the observation point one can see
the predicted shift of the  meeting signal of the particles.
It turns out that for complex systems the whole observation
zone -- of the order of a few fm near the meeting
point -- is filled up by this intensive signal of a strong
interaction scale which we have denoted as "color charge glow".

We emphasize that the new 'color charge glow' effect,  which
was not identified
explicitly in similar studies in the past \cite{MMR98},  is
a manifestation of the non-linear nature of non-Abelian field
dynamics and intimately related to  the color vector rotation
which results in additional field strength from the
previous encounter of color charges moving with velocity $v<1$.
This effect essentially becomes visible in the chromo field
strength and may be not observed in global time-integrated
observables such as the gluon energy and multiplicity
distributions. In this respect, the effect of color charge glow is
as robust as the color rotation itself.

In fact, there is no direct evidence of the Coulomb law for the
interaction between quarks. The leading hypothesis to explain the
observed behavior of quarks is the idea of vacuum gluon fields or
gluon-field fluctuations.  Our considerations do not provide an
alternative to the CGC description of the very initial phase of
ultrarelativistic nucleus-nucleus or proton-nucleus collisions
but should be important at the later stage of the glasma
evolution -- when the CGC is  melting and converting to a QGP with
a considerable amount of quarks and antiquarks -- as well as for
the dynamics of hadronization at the late stage of the QGP
evolution. In this paper, we have not touched upon these topics, but
one should note that estimates based on the instanton vacuum model
indicate that the vacuum field may be much stronger than the
fields generated by the collision of quarks~\cite{mz}. Every open
color in the instanton medium is screened, which indicates the
impossibility of forming gluon fields of noticeable intensity and,
in a physical sense, apparently implies the transformation of
the gluon field quanta in energetically more favorable
configurations.

 With respect to our discussion of the
energy of a non-Abelian system [see
Eq.~(\ref{o3})], we speculate that the signals of the additional
repulsive non-Abelian interaction might leave its traces in the
early development of the collective flow in relativistic
nucleus-nucleus collisions prior to partonic equilibration.

Using the proposed approximation that particles begin to feel the
presence of the non-Abelian charge of a collision partner only
below the distance of $1$ fm, the discussed color configurations
can apparently be considered as a model for a  color
pre-quark/gluon matter. As to a possible formation of some objects
such as a color-flux tube, a larger system similar to that formed in
ultrarlativistic heavy-ion collisions might be considered. Here
new collective effects like the Debye screening come into play,
additionally. However, these effects are beyond the aim of our
study and are the subject of future investigations. Also, an
important "practical" \ conclusion is that, in essence, the correct
assessment for the strengths of chromoelectric and chromomagnetic
fields is obtained just in the approximation of pure
"electrodynamics" \ with enhanced coupling $g^0$. At the same
time, the   experience gained so far \cite{msz} indicates that
such a signal of "color charge glow" \ might manifest itself not
so directly as suggested by the figures above. Subsequent
parton-parton collisions in dense matter -- as produced in
ultrarelativistic Pb+Pb (or Au+Au) collisions -- may wash out this effect.
Accordingly, the study of collisions of light nuclei or proton-nucleus
interactions appears more promising.

\vspace{5mm}
{\bf Acknowledgments} \\[2mm]
We are thankful to Michael Ilgenfritz  and Sergei Nedelko for
illuminating discussions.

\section*{Appendix}
\subsection*{ Time hierarchy of interaction stages}

In the discussion of  a color charge, a color dipole and  two
dipoles we need to compare the description at  different scales.
It is convenient to specify them by the following items:

\begin{enumerate}
\item {\bf The time }$t <T$, $T\leq t<t''$ (recall $t''<0$)\\
The form of the solution for negative times has already
been specified above, for example, the  color charge vectors are
constant up to the scale $t'$
\begin{eqnarray}
\label{35}
&&\widetilde P=\widetilde P_T~,~~t<t_1'\nonumber \\[- .2 cm]
\\[- .25 Cm]
&&\widetilde Q=\widetilde Q_T~,~~t<t_2'~. \nonumber
\end{eqnarray}
Further, up to the time $ t''$,  each charge rotates
relative to the constant vector of the opposite charge particle. Let
us introduce the basis (static) vectors in the color (isotopic)
space, in terms of which it will be convenient to express the
solutions of equations.  The triple orthogonal unit vectors are
associated with the vector $\widetilde P$ in the form:
$$\widetilde Q_T~,~~\widetilde n_{P_T}=\frat{\widetilde P_T \times
\widetilde Q_T}{\sin \theta}~,~~
\widetilde m_{P_T}=\widetilde Q_T \times \widetilde n_{P_T}~,$$
where
$\cos\theta=(\widetilde P_T \widetilde Q_T)$. A similar basis
associated with the vector $ \widetilde Q $ is defined as
$$\widetilde P_T~,~~\widetilde n_{Q_T}=-\widetilde n_{P_T}~,~~
\widetilde m_{Q_T}=\widetilde P_T\times\widetilde n_{Q_T}~.
$$
The solution of Eqs. (\ref{32}) for the times considered
(\ref{35}) can be represented as follows:
\begin{eqnarray}
\label{36} && \widetilde P_{<}=\cos\theta~\widetilde Q_T+ \sin
\theta~\{\cos [\omega(\chi-\chi_1')]~\widetilde
m_{P_T}\nonumber \\&-& \sin[\omega
(\chi-\chi_1')]~\widetilde n_{P_T}\}~,~~
t_1' \leq t<t''~, \nonumber \\[-.2cm]
\\[-.25cm]
&&\widetilde Q_{<}=\cos \theta~\widetilde P_T + \sin\theta~ \{\cos
[\omega (\chi-\chi_2 ')]~\widetilde m_{Q_T} \nonumber \\ &-&
\sin[\omega (\chi-\chi_2 ')]~\widetilde n_{Q_T} \}~,~~
t_2'\leq t<t''~, \nonumber
\end{eqnarray}
where we use the notation $\chi_1'=-\ln|t_1'|$,
$\chi_2'=-\ln|t_2'|$.

\item {\bf The time $ t'' \leq t \leq-t''$} \\
Now consider the asymptotic solutions for large $\chi$ (in the
neighborhood of the meeting point of particles). In this case, one
can  neglect the retardation factors $ \Delta^*$, $ \Delta^{**} $
in Eqs. (\ref{33}), (\ref{34}). Then, for both positive and
negative times these equations describe the  rotation of the color
charge vector with respect to the vector $\widetilde P+\widetilde
Q$ which is conserved. It seems reasonable to approximate the
solutions of Eqs.~(\ref{32}) on the whole semi-axis of the
negative time including the time of the meeting of particles, by
matching the solutions on the $t'$ scale with those on the $t''$
scale,  where the retardation is neglected. The numerical study of
the system (\ref{32}) justifies in general such an approximation.
Thus, at the time  $t''$ the color vectors are
$$\widetilde P'' = \widetilde P_{<}(t'')~,~~
\widetilde Q''=\widetilde Q_{<}(t'')~.$$ Therefore,  the vector
$\widetilde P+\widetilde Q$ is  defined with respect to which
vector of the color charge particles is rotated on the scale $
T''<t<0$. Similarly to the considered case of solutions for large
time $t'$, let us introduce three orthogonal unit vectors
\begin{eqnarray}\widetilde \Omega&=&\frat{\widetilde P''+\widetilde
Q''} {|\widetilde P''+\widetilde Q''|}~,~ \widetilde n_
{P''}\frat{\widetilde P''\times \widetilde \Omega} {|\widetilde
P'' \times \widetilde \Omega |}~,\nonumber \\~~ \widetilde
m_{P''}&=&\widetilde \Omega \times \widetilde n_{P''}~, \nonumber
\end{eqnarray} by means of which the solutions of systems
(\ref{33}) can be expressed as follows
\begin{eqnarray}
\label{37}  \widetilde P_{<} &=& c~\widetilde \Omega+s~\{\cos
[\omega'' (\chi- \chi'')] ~\widetilde m_{P''}\nonumber
\\ &-& \sin[\omega'' (\chi-\chi'')]~\widetilde n_{P''}\}~,
\nonumber \\
[-.2cm]
\\[- .25 cm]
\widetilde Q {<}&=&c~\widetilde \Omega -s~\{\cos
[\omega''(\chi-\chi'')] ~\widetilde m_{P''}\nonumber \\
&-&\sin [\omega''(\chi-\chi'')]~\widetilde n_{P''}\}~,
\nonumber
\end{eqnarray}
\begin{eqnarray}\omega''&=&\omega~|\widetilde P''+\widetilde
Q''|,~~
c=\cos\frat{(\widetilde P'' \widetilde Q'')}{2}~,\nonumber \\
s&=&\sin\frat{(\widetilde P'' \widetilde Q'')}{2}~,
~~\chi''=-\ln |t''|~.\nonumber \end{eqnarray} This solution
describes an infinitely fast rotation of the charges in the
neighborhood of their meeting point, if they are treated  as a
function of time, and we have a rotation with a constant frequency
$\omega''$ in the logarithmic variables $\chi $. To continue
the solution beyond the singularity for positive times, we assume
that in the time interval $-t_{\mbox{\small
{min}}}<t<t_{\mbox{\small {min}}}$, the particle charge vector
does not change. The time scale $t_{\mbox{\small {min}}}$ is of
artificial character and can be chosen arbitrarily small, in
principle. Without retardation  the solution of Eq. (\ref{34}) at
$t>t_{\mbox{\small {min}}}$ can be written as
\begin{eqnarray}
\label{38} \widetilde P_ {>}&=&c~\widetilde \Omega
+s~\{\cos[\omega''(\chi-\chi^*)]~\widetilde m_ {P''}
\nonumber \\ &+&
\sin[\omega''(\chi-\chi^*)]~\widetilde n_{P''}\}~, \nonumber \\
[- .2 cm]
\\[-.25cm]
\widetilde Q_{>}&=&c~\widetilde \Omega-s~\{\cos
[\omega''(\chi-\chi^{**})] ~\widetilde m_{P''} \nonumber
\\ &+&\sin[\omega''(\chi-\chi^{**})]~\widetilde n_{P''}\}~,
\nonumber
\end{eqnarray}
where $\chi=-\ln t$, $\chi^*$, $\chi^{**}$ are arbitrary
phases to be  defined by the known values of the charge vectors on
the left end of the segment $-t_{\mbox{\small {min}}}$
\begin{eqnarray}\widetilde P_{<}(-t_{\mbox{\small {min}}}) &=& \widetilde
P_{>}(t_{\mbox{\small {min}}})~,~\nonumber \\~ \widetilde
Q_{<}(-t_{\mbox{\small{min}}})&=& \widetilde
Q_{>}(t_{\mbox{\small{min}}})~.\nonumber \end{eqnarray}
 Comparing Eq.~(\ref{37}) and (\ref{38}), one can obtain relations
between the phase $\chi''$ and phases $\chi^*$ and
$\chi^{**}$. It is seen that it is convenient to put the time $
t_{\mbox {\small {min}}} $ in such a way that
$\sin[\omega''(\chi_{\mbox{\small{min}}}-\chi'')]=0$, with
$\chi_{\mbox{\small{min}}}=-\ln t_{\mbox{\small{min}}}$; i.e.
$\chi_{\mbox{\small{min}}}=2\pi~n/\omega''+\chi''$, where
$n$ is an integer number. Then one can put
$\chi^*=\chi^{**}=\chi''$. As a result, the expression
for  positive times will not be overloaded by formal phase shifts
and at the same time an arbitrarily small time scale
$t_{\mbox{\small{min}}}$ can be chosen. Thus, the solution for the
time scale $t_{\mbox{\small{min}}}<t<|t''|$ has been obtained. As
a result, we can see that \begin{eqnarray}\widetilde
P_{<}(t'')&=&\widetilde P_{>}(|t''|)=\widetilde P''~,~\nonumber \\
\widetilde Q_{<}(t'')&=&\widetilde Q_{>}(|t''|)=\widetilde Q''~,
\nonumber\end{eqnarray} i.e. in this approximation the phase shift
is not observed, and the color charge vector at the exit from the
scale $t''$ coincides with that at its entrance.

\item {\bf The time:} $-t''<t\leq t_{\mbox{\small{out}}}$,
$t_{\mbox{\small{out}}}<t$\\
We  continue the approximate solution to larger times in such a way
that a change of the color rotation regime on the scale
$t_{\mbox{\small{out}}}\sim|t'|$ should occur symmetrically with
respect to the negative time and the rotation should stop when the
scale $t\sim |T|$ is reached. This solution for the positive time
can be written in the form [a change of the
sign in this system in comparison with Eq.~(\ref{36}) should be mentioned]
\begin{eqnarray}
\label{39}  \widetilde P_{>} &=& \cos \theta_o~\widetilde Q_o +
\sin \theta_o~\{\cos[\omega(\chi-\chi_o^*)]~ \widetilde
m_{P_o} \nonumber \\ &+& \sin[\omega
(\chi-\chi_o^*)]~\widetilde n_{P_o}\}~,
~~|t''|\leq t<|t'|~, \nonumber \\[-.1cm]
\\[-.2cm]
\widetilde Q_{>} &=& \cos \theta_o~\widetilde P_o + \sin
\theta_o~\{\cos[\omega(\chi-\chi_o^{**})]~\widetilde
m_{Q_o}\nonumber \\ &-&
\sin[\omega(\chi-\chi_o^{**})]~\widetilde n_{Q_o}\}~,
\nonumber
\end{eqnarray}
with yet unknown basis vectors
\begin{eqnarray}\widetilde
Q_o~,~~\widetilde n_{P_o}&=&\frat{\widetilde P_o \times \widetilde
Q_o}{\sin \theta_o}~,~~\widetilde m_{P_o}= \widetilde Q_o \times
\widetilde n_{P_o}~,~~\nonumber \\ \widetilde P_o~,~~ \widetilde
n_{Q_o}&=&-\widetilde n_{P_o}~,~~ \widetilde m_{Q_o}=\widetilde
P_o \times \widetilde n_{Q_o}~,\nonumber
\end{eqnarray}
on which the solution is spanned and with the corresponding phases
$\chi_o^*$, $\chi_o^{**}$. Here $\cos \theta_o =(\widetilde
P_o \widetilde Q_o)$. This information should be restored using
the available $\widetilde P_{>}(|t''|)=\widetilde P''$,
$\widetilde Q_{>}(|t''|)=\widetilde Q''$. The vector product
$\widetilde P''\times \widetilde Q''$ gives another condition in
addition to the relations (\ref{39}). Applying the vector algebra
rules, these relations can be presented in the matrix form
\begin{eqnarray}
\label{40} &&a_{11}~\widetilde P_o+a_{12}~\widetilde
Q_o+a_{13}~\widetilde
n_{P_o}=\widetilde P''~,\nonumber\\
&&a_{21}~\widetilde P_o+a_{22}~\widetilde Q_o+a_{23}~\widetilde
n_{P_o}=\widetilde Q''~,\\
&&a_{31}~\widetilde P_o+a_{32}~\widetilde Q_o+a_{33}~\widetilde n_{P_o}=
\widetilde P'' \times \widetilde Q''~, \nonumber
\end{eqnarray}
with the coefficients
\begin{eqnarray}
&&a_{11}=c_1~,~~a_{12}=c(1-c_1)~,~~a_{13}=s_1~, \nonumber \\
&&a_{21}=c(1-c_2)~,~~a_{22}=c_2~,~~a_{23}=-s_2~, \nonumber \\
&&a_{31}=(\alpha-\beta c)/s~,~~a_{32}=(\beta-\alpha c)/s~,
~~a_{33}=\gamma s~, \nonumber
\end{eqnarray}
where the following notation is used:
\begin{eqnarray}
\alpha&=&-c_2s_1-c(1-c_1)~s_2~,~\nonumber
\\ \beta&=&c_1s_2+c(1-c_2)~s_1,\nonumber
\\ \gamma&=&c_1 c_2-c^2 (1-c_1) (1-c_2)~, \nonumber \\
c&=&\cos \theta_o=(\widetilde P_o \widetilde Q_o),\nonumber \\~~
s&=&\sin \theta_o~,~~c_1=\cos\theta_o^*~,\nonumber \\~~s_1&=&\sin
\theta_o^*~,~~
\theta_o^*=\omega(\chi''-\chi_o^*)~, \nonumber \\
c_2&=&\cos \theta_o^{**}~,~~s_2=\sin \theta_o^{**}~,~~
\theta_o^{**}=\omega(\chi''-\chi_o^{**})~. \nonumber
\end{eqnarray}
Now, if all the phases are known, then  one can find $ \widetilde
P_o $, $ \widetilde Q_o $ using the inverse  matrix $A^{-1}$ .

In order to determine the phases, we use an important relation for
the scalar product of vectors $\widetilde P''$ and $\widetilde
Q''$
\begin{eqnarray}
\label{41} && (1-c_1) (1-c_2)~c^3+(c_1+c_2-c_1 c_2)~c-s_1 s_2
\nonumber \\ &=& (\widetilde P'' \widetilde Q'')~.
\end{eqnarray}

\end{enumerate}

Numerical analysis with the coefficients $c_1$, $c_2$ ($s_1$, $s_2$)
shows that the resulting cubic equations for the cosine of the
angle between the vectors $\widetilde P_o$ and $\widetilde Q_o$
has one real and two complex conjugate roots. The real root does
not always satisfy the restriction $|c|<1 $, i.e. in general all of
the coefficients $c_1$, $c_2$, ($s_1$, $s_2$) and $c$ should  be
consistent. The boundaries of the acceptable region are defined by
setting $c=\pm 1$, then
$$\pm(1\mp s_1 s_2)=(\widetilde P'' \widetilde Q'')~,$$
In particular, for the parallel and anti-parallel vectors
$\widetilde P''$ and $\widetilde Q''$ we have $s_1=s_2=0$.
For an approximate solution we can restrict ourselves to a
particular case $s_1=s_2=0$ just as we did in choosing the
phase $\chi_{\mbox{\small{min}}}$. Then for the
phase one can get
$$\omega(\chi''-\chi_o^*)=0+2\pi~n~,~~
\omega (\chi''-\chi_o^*)=\pi+2\pi~n~,$$ where $n$ is an
integer. Similar relations hold for the phase $\chi_o^{**}$.
The time scale, when the rotation around the constant  charge
vector of the particle-partner stops, is defined by the condition
$t_{\mbox{\small {out}}}=e^{-\chi_o^*}$. The analyzed cases
show that for the considered particle velocities (with reasonable
accuracy) will fall at the scale of $| t'|$, if  the phase is
chosen as $\chi_o^*=\chi_o^{**}=\chi''-\pi/\omega$. This
regime is approximately applied when $m/{\cal E}<10^{-1.5}$ and
it gets better with an increase in energy reaching the required
scale. In principle, nothing prevents  installing the time
$t_{\mbox{\small{out}}}$ also for moderate relativistic
velocities, because at this scale $\sim t'$  the change in color
charges is insignificant, as compared to the scale $t''$. With
this choice of the phase, Eq.~(\ref{41}) takes the form
$$4~c^3-3~c=(\widetilde P''\widetilde Q'')~.$$
This equation has one real root which always satisfies the
condition $|c|<1$, and two conjugated imaginary roots. Now with
the known coefficients $c_1$, $c_2$, ($s_1$, $s_2$) and $c$, one
can find the vectors of the particles that define the basic three
vectors which are spanned over the solutions for positive time
scales $t_{\mbox{\small{out}}}$ and $|T|$. The  equation set
(\ref{40}) for  particular cases interested takes the form
\begin{eqnarray}
&&\!\!\!-\widetilde P_o+2 c~\widetilde Q_o=\widetilde P''~,\nonumber\\
&&2c~\widetilde P_o-\widetilde Q_o=\widetilde Q''~. \nonumber
\end{eqnarray}
>From here we get the solutions
\begin{eqnarray}
&&\widetilde P_o=\frat{1}{4c^2-1}~\widetilde P''+\frat{2c}{4c^2-
1}~\widetilde
Q''~, \nonumber \\
&&\widetilde Q_o=\frat{2c}{4c^2-1}~\widetilde P''+\frat{1}{4c^2-
1}~\widetilde
Q''~. \nonumber
\end{eqnarray}
The explicit expressions provide  an approximate solution of
Eq.~(\ref{32}) continuous in time which can be applied to the
entire time axis.

\subsection*{Color charge and dipole}
As before we assume that  particles begin to feel the presence of
the third-particle color charge when approaching the distance $D$
estimated as $1$ fm. The signal of the presence of the charge
$\widetilde P$ arrives at the second particle at time $t_2'$, and
at the third one at $t_3'=t_2'+t_3$. We ignore the
 time difference between $t_2'$ and $t_3'$ (due to the Lorentz
contraction), i.e. the charges of the second and third particles
do not change up to the time $t_2'$. With the same degree of
accuracy, the signals from the second and third charges will come
to the first particle at the time $t_1'$. Until this point the
first particle charge $\widetilde P$ does not change in time. Then
the charges rotate with respect to the constant charge vector,
which the particles had at the entrance in the interaction zone.
For these times  the compatibility conditions (\ref{39}) become
\begin{eqnarray}
\label{44}
&&\dot{\widetilde P}=\omega~\left[\frat{1}{|t|}-\frat{1}{|t-t_3|}\right]~
\widetilde Q_T \times \widetilde P~, \nonumber\\
&&\dot{\widetilde Q}_2=\omega~\frat{1}{|t|}\widetilde P_T \times
\widetilde Q_2
-\omega_\|~\frat{1}{t_3}\widetilde Q_T \times \widetilde Q_2~, \\
&&\dot {\widetilde Q}_3=\omega~\frat{1}{|t-t_3|}
\widetilde P_T \times \widetilde Q_3-\omega_\|~\frat{1}{t_3}
\widetilde Q_T \times \widetilde Q_3~, \nonumber
\end{eqnarray}
with $\omega_\|=\frat{\alpha_g~(1-w^2)}{v+w}$. From
this system one can conclude that due to the factor of $\omega_\|$
the charges of $\widetilde Q_2$ and $\widetilde Q_3$  can be
considered as mutually opposite well away from the meeting point
(as well as at the entrance to the interaction zone), in particular,
until the scale $t'$, the dynamics of color charges is described
by a simplified system of the two equations
\begin{eqnarray}
\label{45}
&&\dot{\widetilde P}=\omega~\left[\frat{1}{|t|}-\frat{1}{|t-t_3|}\right]~
\widetilde Q_T \times \widetilde P~, \nonumber\\[-.2cm]
\\[-.25cm]
&&\dot{\widetilde Q}=\omega~\frat{1}{|t|}
\widetilde P_T \times \widetilde Q~, \nonumber
\end{eqnarray}
$\widetilde Q_2=\widetilde Q$, $\widetilde Q_3=-\widetilde Q$.
Comparing this set of equations with Eqs. (\ref{32}) we get a
solution in the form
\begin{eqnarray}
\widetilde P_{<}&=&\cos\theta~\widetilde Q_T +
\sin\theta~\{\cos[\omega(\eta-\eta_1')]~\widetilde
m_{P_T}\nonumber \\ &-& \sin[\omega(\eta-\eta_1')]~\widetilde
n_{P_T}\}~,~~t_1'\leq t<t''~,
\nonumber \\
\widetilde Q_{<}&=&\cos\theta~\widetilde
P_T+\sin\theta~\{\cos[\omega(\chi- \chi_2')]~ \widetilde
m_{Q_T} \nonumber \\
&-& \sin[\omega(\chi-\chi_2')]~\widetilde
n_{Q_T}\}~,~~t_2'\leq t<t''~, \nonumber
\end{eqnarray}
where
$$\eta=\chi-\psi~,~~\chi=-\ln|t|~,~~\psi=-\ln|t-t_3|~,$$
with the initial data on the scale $t'$. In this case, the  dipole
charges act on the color charge of the first particle weaker than
a single color charge because the contributions compensate  each
other. This scheme is equivalent to item (1) of the previous
section, and describes the behavior of charges for times $t <T$,
$T\leq t <t''$.

As in the case of the two color charges let us take  as an
acceptable approximation to the exact solution the matching of the
solution on the  scale $t''$, which neglects the retardation, and
the above solutions for the time scale  larger than $ t '$. In the
meeting area  the charges of the first and second particles are
described by the familiar equations
\begin{eqnarray}
\label{46}
&&\dot{\widetilde P}=\omega~\frat{1}{|t|}~\widetilde Q_2
\times\widetilde P~,\nonumber\\[-.2cm]
\\[-.25cm]
&&\dot{\widetilde Q}_2=\omega~\frat{1}{|t|}
\widetilde P \times \widetilde Q_2~. \nonumber
\end{eqnarray}
In contrast to the singular behavior dictated by this
system, the charge of the third particle obeys
$$\widetilde Q_3\simeq\omega~\frat{1}{t_3}~\widetilde P_T\times\widetilde
Q_3~.$$ In many applications the time scale $t''$ is so small that
in this interval the color vector $\widetilde Q_3$ can even be taken
as a constant. As was mentioned in the section devoted to the two
color charges, after the singularity point for a positive time
$t=|t''|$  the charges take the same position in the color space as
at the entrance to the singularity zone. Therefore, the continuation
to larger positive times  used for the two color charges holds valid
in the color-dipole case, the color charge dynamics being described
by the simplified system of equations (\ref{45}). The initial data
are defined by the obvious matching of the condition with a singular
solution at the time $t=|t''|$, which we do not present here. It is
clear that the same considerations of the behavior of color charges
can be applied to the meeting point of the first charge with the
third particle on the scale of $t''$ in the neighborhood of $t_3$.
\begin{eqnarray}
\label{47}
&&\dot{\widetilde P}=\omega~\frat{1}{|t-t_3|}~
\widetilde Q_3\times\widetilde P~, \nonumber\\[-.2cm]
\\[-.25cm]
&&\dot{\widetilde Q}_3=\omega~\frat{1}{|t-t_3|}
\widetilde P \times \widetilde Q_3~.\nonumber
\end{eqnarray}
In its turn, the charge vector of the second particle $\widetilde
Q_2$ can be considered in this segment as constant. As in item (ii)
of the previous section for the times $t''\leq t\leq-t''$, here
for the times $ t'' \leq t-t_3 \leq-t''$ one  should construct the
appropriate three  basic vectors based on the continuous vector
$\widetilde P(t_3+t'')+\widetilde Q_3(t_3+t'')$ and introduce an
additional time scale $t_{\mbox{\small{min}}}$ which is defined by
the relation
$\chi_{\mbox{\small{min}}}=2\pi~n/\omega_{t_3+t''}+\chi''$,
$\Omega_{t_3+t''}=|\widetilde P(t_3+t'')+\widetilde
Q_3(t_3+t'')|$. In this way we obtain  solutions analogous to
those in item (ii) but for the case of two color charges  at the
meeting points of the particles 1-2 and 1-3.

The continuation of the solution to larger times should be carried
out by analogy with item (iii) of the previous section, i.e., first, to
construct the description of  charges on the scale $-t'$, then on
the scale of $t_3+t'$ and so on, up to the scale $t_3+t''$ where
there is a meeting of the first and third particle and the
behavior of charges is singular. But such a meticulous description
apparently is not needed if  approximate solutions are
considered. So we just accept that the passage to the regime of
rotation around constant vectors of color charges of
particles-partners at the exit from the interaction zone occurs
somewhere on the scale $t_3$. Such a solution of (\ref{45}) is as
follows~:
\begin{eqnarray}
\widetilde P_{>}&=&\cos\theta_o~\widetilde Q_o
+\sin\theta_o~\{\cos[\omega(\eta-\eta_o^*)]~\widetilde m_{P_o}
\nonumber \\ &+&\sin[\omega(\eta-\eta_o^*)]~\widetilde
n_{P_o}\}~,~~|t''|\leq
t<t_{\mbox{out}}~,\nonumber\\
\widetilde Q_{>}&=&\cos\theta_o~\widetilde P_o
+\sin\theta_o~\{\cos[\omega(\chi-\chi_o^{**})]~\widetilde
m_{Q_o}\nonumber \\ &-&
\sin[\omega(\chi-\chi_o^{**})]~\widetilde n_{Q_o}\} ~,
\nonumber
\end{eqnarray}
where as in the previous section, one should determine the basic three
vectors and phases using available information  on the scales $t''$,
i.e.,
$\widetilde P''$, $ \widetilde Q''$. The corresponding systems of
equations have the form (\ref{40}), (\ref{41}), where the variable
substitution  $ \chi \to \eta $ should be made for variables
related with the vector of the first particle charge $ \widetilde
P $. As was noted in the previous section, the easiest version of
the inverse problem of reconstruction of the basis vectors and
phases  would be appropriate if this choice of variables gives $
s_1 = s_2 = 0 $. An analysis shows that for the problem of the
particle and the dipole one can take
$$\omega(\eta''-\eta_o^*)=2\pi~,~~\omega(\chi''-
\chi_o^{**})=\pi~.$$
With this choice of phases, going on to large positive times
occurs somewhere on the scale of $t_3$ for the first particle and
on the scale $t_2'$ for the second  one
\begin{eqnarray}t_{o1}&=&\frat{t_3}{1-e^{-
x_{o1}}}~,~~x_{o1}=\frat{2\pi}{\omega}-\eta''~~
\nonumber
\\ t_{o2}&=&e^{-x_{o2}}~,~x_{o2}=\chi''-\frat{\pi}{\omega}~.\nonumber
\end{eqnarray}
 As mentioned above, there is no sense in complicating
the task by better matching the transition regime for
asymptotically large times. Now Eq. (\ref{41}) becomes ($c_1=1$, $c_2=-
1$)
$$c=(\widetilde P'' \widetilde Q'')~.$$
The system of equations (\ref{40}) in which we are particularly
interested takes the form
\begin{eqnarray}
&& \widetilde P_o = \widetilde P''~, \nonumber \\
&& 2c~\widetilde P_o - \widetilde Q_o = \widetilde Q'' ~. \nonumber
\end{eqnarray}
In this way we complete the description referred to the item (iii) for
the problem of two colors charges. The whole procedure to obtain
approximate solutions for the  system of compatibility equations
(\ref{43}) is reduced to the description of the behavior of two
basic charges, since the partner charge in a dipole pair can be
considered as adjusted. The passage for a short time on the scale
$t''$ to a singular rotation regime is not accompanied by a
change in phase.

We have missed some interesting effects of the arrival of the
signal to the particle-partner in the dipole from the meeting
point of two other particles. In  Fig.~\ref{f8}, the marked point
$12$ corresponds to the light signal  coming from the meeting of
the first and second charges to the third color charge. In these
times on the scale $t''$ the charge of the third particle is
described by the equation
\begin{equation}
\label{48}
\widetilde Q_3 \simeq\omega~\frat{1}{t_3}~
\widetilde P\times\widetilde Q_3~,
\end{equation}
where the charge of the first particle $ \widetilde P $ is rapidly
changing in a singular manner. Unfortunately, it is difficult to
derive analytical expressions describing the behavior of the color
charge of the third particle, and in this paper we simply ignore
this important, but short,  episode. It is also important to note
in turn  that  a signal about events that happened to the third
charge prior to the collision of the first charge with the third
particle reaches in time the first particle. Then, a signal will
come to the third particle even before the meeting of the first
and the third particles, and so on. There is some danger that we
are not able to control the behavior of charges at the second
meeting point because one should  trace the ladder of events up to
the meeting point (we have not depicted in Fig.~\ref{f8} an
appropriate sequence of signals similar to those shown in Fig.~\ref{f1}).
However, it is noteworthy  that these processes should
not change too much the color charge phase since the first signal
comes on the scale $t_3\sim m/{\cal E}$, and the second one occurs
on the scale $t_3\times t''\sim m^5/{\cal E}^5$, while the merging
of solutions takes place on the scale $t''$.

\subsection*{Two color dipoles}
We discard a detailed prescription  for obtaining approximate
solutions,  as was done in the previous Sections. Let us write down
directly the simplified system of equations which allows us to describe
the behavior of color charges for large times,
\begin{eqnarray}
\label{51}
&&\dot{\widetilde P}=\omega~\left[\frat{1}{|t|}-\frat{1}{|t-t_3|}\right]~
\widetilde Q_T \times \widetilde P~, \nonumber\\[-.2 cm]
\\[-.25cm]
&&\dot{\widetilde Q}=\omega~\left[\frat{1}{|t|}-\frat{1}{|t-t_4|}\right]
\widetilde P_T \times \widetilde Q~, \nonumber
\end{eqnarray}
where $\widetilde P_1 = \widetilde P$, $\widetilde P_4=-\widetilde P$,
$\widetilde Q_2=\widetilde Q$, $\widetilde Q_3=-\widetilde Q$.
One should specify that under
construction of approximate solutions there are additional basic
three vectors by means of which solutions at the meeting point of
particles are  built. One should also introduce the appropriate
set of times $t_{\mbox{\small{min}}}$, where formally singular
solutions are matched. On a time scale $t'$ the solution of the
system (\ref{51}) has the form
\begin{eqnarray}
 \widetilde P_{<}&=&\cos\theta~\widetilde Q_T+\sin\theta~
\{\cos[\omega(\eta-\eta_1')]~\widetilde m_{P_T} \nonumber \\ &-&
\sin[\omega(\eta-\eta_1')]~\widetilde n_{P_T}\}~,~~t_1'\leq
t<t''~,
\nonumber \\
\widetilde Q_{<}&=&\cos\theta~\widetilde P_T+\sin\theta~
\{\cos[\omega(\zeta-\zeta_2')]~\widetilde m_{Q_T} \nonumber \\ &-&
\sin[\omega(\zeta-\zeta_2')]~\widetilde n_{Q_T}\}~,~~t_2'\leq
t<t''~, \nonumber
\end{eqnarray}
where
 \begin{eqnarray}
 \eta&=&\chi-\psi~,~~\zeta=\chi-\xi~,~~
\chi=-\ln|t|~,~\nonumber
\\ \psi&=&-\ln|t-t_3|~,~~~~\xi=-\ln|t-t_4|~,\end{eqnarray} with the
initial data taken on the scale $t'$. The passage of approximate
solutions from the scale $|t''|$ to large positive times is
described by the solution of the form
\begin{eqnarray}
\widetilde P_{>}&=&\cos\theta_o~\widetilde Q_o
+\sin\theta_o~\{\cos[\omega(\eta-\eta_o^*)]~\widetilde m_{P_o}
\nonumber \\ &+&\sin[\omega(\eta-\eta_o^*)]~\widetilde
n_{P_o}\}~,~~|t''|
\leq t<t_{\mbox{out}}~,\nonumber \\
\widetilde Q_{>}&=&\cos\theta_o~\widetilde P_o
+\sin\theta_o~\{\cos[\omega(\zeta-\zeta_o^{**})]~\widetilde
m_{Q_o} \nonumber \\
&-& \sin[\omega(\zeta-\zeta_o^{**})]~\widetilde n_{Q_o}\}~,
\nonumber
\end{eqnarray}
where the three basic vectors and phase are determined by the
 conditions $s_1=s_2=0 $ discussed above. Under these conditions
one  can avoid solutions of the  complicated inverse problem
of the restoration of basic triple vectors using the initial data
on the scale $ t'' $. In the case of two dipoles one can take
$$\omega(\eta''-\eta_o^*)=2\pi~,~~\omega(\zeta''-\zeta_o^{**})=2\pi~.$$
With this choice of the phase the passage to large positive times
for the first and fourth particle occurs somewhere on the scale
$t_4$ and  on the scale $t_3$ for the second and third particle
\begin{eqnarray}
t_{o1}&=&\frat{t_3}{1-e^{-
x_{o1}}}~,~~x_{o1}=\frat{2\pi}{\omega}-
\eta''~,\nonumber
\\~~t_{o2}&=&\frat{t_4}{1-e^{-x_{o2}}}~,~~x_{o2}=\frat{2\pi}{\omega}-
\zeta''~.\end{eqnarray}
 Accordingly,  in the case of two dipoles the
passage to large positive times is more consistent than in the
particle-dipole case. Now Eq. (\ref{41}) becomes ($c_1=1$, $c_2=1$)
$$c=(\widetilde P''\widetilde Q'')~.$$
For a particular case considered the system  (\ref{40}) is
reduced  to
$$\widetilde P_o=\widetilde P''~,~~\widetilde Q_o=\widetilde Q''~.$$
Thus, the approximate solution on the entire time axis has been
constructed.

\end{document}